\documentclass[longauth]{aa}
\usepackage{graphicx}
\usepackage{txfonts}
\usepackage{booktabs}
\usepackage{verbatim,multirow,multicol}
\usepackage{hhline}
\usepackage{soul}
\setstcolor{red}
\setstcolor{black}
\usepackage{color}
\usepackage{tablefootnote}
\usepackage{url}

\newcommand{\Gaia}{\textit{Gaia}~}

\begin{document}

   \title{Constraints on (2060) Chiron's size, shape, and surrounding material from the November 2018 and September 2019 stellar occultations}
   
   \titlerunning{Constraints on (2060) Chiron's size, shape and surroundings from stellar occultations}
   

    \authorrunning{F. Braga-Ribas, C. L. Pereira, B. Sicardy et al.}
    
   \author{F. Braga-Ribas\inst{1,2,3}\fnmsep\thanks{\email{fribas@utfpr.edu.br}}
   \and C. L. Pereira \inst{4,3,1}
   \and B. Sicardy\inst{2} 
   \and J. L. Ortiz \inst{5} 
   \and J. Desmars \inst{6,7} 
   \and A. Sickafoose \inst{8} 
   \and M. Emilio \inst{9}
   \and B. Morgado \inst{10, 3} 
   \and G. Margoti \inst{1, 3} 
   \and F. L. Rommel \inst{1,3} 
   \and J. I. B. Camargo\inst{4,3} 
   \and M. Assafin\inst{10,3} 
   \and R. Vieira-Martins \inst{4,3} 
   \and A. R. Gomes-J\'unior \inst{11,3}
   \and P. Santos-Sanz\inst{5} 
   \and N. Morales\inst{5} 
   \and M. Kretlow\inst{5,12} 
   \and J. Lecacheux\inst{2} 
   \and F. Colas\inst{7} 
   \and R. Boninsegna\inst{12} 
   \and O. Schreurs\inst{13,14} 
   \and J. L. Dauvergne\inst{15} 
   \and E. Fernandez\inst{13,14} 
   \and H. J. van Heerden\inst{16} 
   \and H. González\inst{17} 
   \and D. Bihel\inst{18}
   \and F. Jankowsky\inst{19}
   }

\institute
{Federal University of Technology - Paraná (PPGFA/UTFPR-Curitiba), Rua Sete de Setembro, 3165, Curitiba, PR, Brazil,
\and 
LESIA, Observatoire de Paris, Université PSL, CNRS, Sorbonne Université, Univ. Paris Diderot, Sorbonne Paris Cité, 5 place Jules Janssen, 92195 Meudon, France,
\and
Laboratório Interinstitucional de e-Astronomia - LIneA \& INCT do e-Universo, Brazil,
\and
Observatório Nacional/MCTIC, R. General José Cristino 77, Bairro Imperial de São Cristóvão, Rio de Janeiro, RJ, Brazil,  
\and
Instituto de Astrofísica de Andalucía, IAA-CSIC, Glorieta de la Astronomía s/n, 18008 Granada, Spain, 
\and
Institut Polytechnique des Sciences Avancées IPSA, 63 boulevard de Brandebourg, F-94200 Ivry-sur-Seine, France,
\and
Institut de Mécanique Céleste et de Calcul des Éphémérides, IMCCE, Observatoire de Paris, PSL Research University, CNRS,Sorbonne Universités, UPMC Univ Paris 06, Univ. Lille, 77 Av. Denfert-Rochereau, F-75014 Paris, France,
\and
Planetary Science Institute, 1700 East Fort Lowell Road, Suite 106, Tucson, AZ 87519
\and
Universidade Estadual de Ponta Grossa, O.A. - DEGEO, Av. Carlos Cavalcanti, 4748, Ponta Grossa 84030-900, PR, Brazil, 
\and
Observatório do Valongo/UFRJ, Ladeira Pedro Antônio 43, Rio de Janeiro, RJ, Brazil,
\and
Institute of Physics, Federal University of Uberlândia, Uberlândia, MG, Brazil,
\and
European Asteroidal Occultation Network (EAON)
\and
International Occultation Timing Association - European Section (IOTA-ES),
\and
Observatoire de Nandrin, Société Astronomique de Liège, Avenue des Platanes 17, B 4000 Liège (Cointe),
Belgium,
\and
Ciel et Espace, Paris, France,
\and
University of the Free State, 205, Nelson Mandela Dr, Park West, Bloemfontein, 9301, South Africa,
\and
Observatorio astronómico de Forcarei OAF 36556 Forcarei Pontevedra, Spain,
\and
Planetarium Ludiver, 1700, rue de la Libértion, Tonneville, 50460 La Hauge, France,
\and 
Landessternwarte, Universit\"at Heidelberg, K\"onigstuhl, D-69117 Heidelberg, Germany.
}

\date{Received 02/05/2023; accepted 15/06/2023}

\abstract
    { After the discovery of rings around the largest known Centaur object, (10199) Chariklo, we carried out observation campaigns of stellar occultations produced by the second-largest known Centaur object, (2060) Chiron, to better characterize its physical properties and presence of material on its surroundings.}
    { We aim to provide constraints on (2060) Chiron's shape for the first time using stellar occultations. We  investigate the detectability of material previously observed in its vicinity using the 2018 occultation data obtained from South Africa Astronomical Observatory (SAAO).} 
    { We predicted and successfully observed two stellar occultations by Chiron. These observations were used to constrain its size and shape by fitting elliptical limbs with equivalent surface radii in agreement with radiometric measurements. We also obtained the properties of the material observed in 2011 with the same technique used to derive Chariklo's ring properties in our previous works, used to obtain limits on the detection of secondary events in our 2018 observation.}  
   {Constraints on the (2060) Chiron shape are reported for the first time. Assuming an equivalent radius of $R_{\rm equiv}= 105^{+6}_{-7}\,\rm{km,}$ we obtained a semi-major axis of $a= 126 \pm 22\,\rm{km}$. Considering Chiron's true rotational light curve amplitude and assuming it has a Jacobi equilibrium shape, we were able to derive a 3D shape with a semi-axis of $a=126\pm22$~km, $b=109\pm19$~km, and $c=68\pm13$~km, implying in a volume-equivalent radius of R$_{vol}$= 98~$\pm$~17~km.
   We determined the physical properties of the 2011 secondary events around Chiron, which may then be directly compared with those of Chariklo rings, as the same method was used. Data obtained from SAAO in 2018 do not show unambiguous evidence of the proposed rings, mainly due to the large sampling time. Meanwhile, we discarded the possible presence of a permanent ring similar to (10199) Chariklo's C1R in optical depth and extension. }
   {Using the first multi-chord stellar occultation by (2060) Chiron and considering it to have a Jacobi equilibrium shape, we derived its 3D shape, implying a density of $1,119~\pm~4~\rm{kg~m}^{-3}$. 
   New observations of a stellar occultation by (2060) Chiron are needed to further investigate  the material's properties around Chiron, such as the occultation predicted for September 10, 2023.}
   

\maketitle


\section{Introduction}
 
 After discovering rings around the Centaur object (10199) Chariklo, it was proposed that this feature may be common among distant small Solar System objects \citep{Braga-Ribas2014}. The centaur object (2060) Chiron is the second largest object of its kind. It has an equivalent diameter of 210$^{+11}_{-14}$~km (model dependent), measured using data from the space telescopes \textit{Spitzer} and \textit{Herschel}, as well as from the Atacama Large Millimeter Array (ALMA) \citep{Lellouch2017}. %
 
 Fundamental properties such as size and shape for objects of this kind can be obtained from Earth-based observations using stellar occultations \citep{Ortiz2020} \footnote{An updated list of the detected stellar occultations by outer Solar System objects that we are aware of can be found at \url{http://occultations.ct.utfpr.edu.br/results} \citep{Braga-Ribas2019}}. 
 High-cadence observations in terms of time and high signal-to-noise ratio (S/N) allow for detecting or setting upper limits to the presence of material around the occulting body, for instance, on dust shells, jets, or rings \citep{elliot1995, Ortiz2020}. 
 
Secondary events on previous stellar occultation by (2060) Chiron have been reported for events in 1993, 1994, and 2011
\citep{bus1996, elliot1995, Ruprecht2015}. They were all interpreted as dust or jets coming from Chiron's surface, as it has presented periods of cometary activity \citep{Tholen1988, Belskaya2010}. In 1989, \cite{Meech89} were the first to detect a comma around Chiron. Later, \cite{Luu90} analyzed 1988 to 1990 data,  further investigating Chiron's activity. As far as we know, no non-gravitational acceleration has been detected so far.

The November 29, 2011 occultation was observed using the Infrared Telescope Facility (IRTF), which detected the main body event, and from Faulkes Telescope North (FTN), which had a much higher cadence and S/N, but with no detection of the main body \citep[][]{Ruprecht2015}. The latter data presented two pairs of sharp secondary events that resembled those observed around Chariklo in 2013. These data were further analyzed by \citet{Sickafoose2020}.

\citet{Ortiz2015} analyzed the reported absolute magnitude variation and the decrease of the amplitude of its rotation light curve, together with  the secondary events relative to Chiron positions detected thus far, to argue that Chiron may also have a ring system. The final word on the presence of rings around Chiron depends on new observations of a stellar occultation with high cadence and S/N or on a visit of a spacecraft such as $Centaurus$ \citep{Singer2019}.
 
In this work, we presented the first multi-chord stellar occultation by Chiron, observed on September 8, 2019, from four different sites. It gives further constraints on Chiron's size and shape. We also present a single chord detection observed from South Africa Astronomical Observatory (SAAO) on November 28, 2018, setting an upper limit on the detection of secondary events.

\section{Stellar occultation of September 08, 2019}\label{2019occ}

This stellar occultation event was predicted by the Lucky-Star project\footnote{\url{https://lesia.obspm.fr/lucky-star/}} using the Numerical Integration of the Motion of an Asteroid (NIMA), as described in \citet{desmars2015}, and the second \Gaia Data Release (DR2) catalog \citep{gaia18}. The prediction path was updated using the astrometric position obtained at Pico dos Dias and Calar Alto Observatories and the astrometric position derived from the November 2018 occultation event (see Sect. \ref{2018occ}). 

The occulted star has the number 2741744913738549760 on the third \Gaia Data Release (DR3) with a G magnitude of 16.52. The event had a shadow velocity of 22.8 $\rm{km\,s}^{-1}$, relative to the geocenter. The position of the star from the \Gaia DR3 catalog was propagated using parallax and proper motion to the event date, resulting in the following geocentric position at the epoch:
\vspace{-0.5cm}
\begin{center}
$$ \alpha = 00h~10m~12s.74303 \pm 0.1600~{\rm mas,} $$ 
\vspace{-5mm}
$$ \delta = +04^\circ~37'~04''.9141 \pm 0.2824~{\rm mas}, $$
\end{center}
where mas stands for milliarcsec.

\begin{figure}[!h]
    \centering
    \includegraphics[width=\hsize]{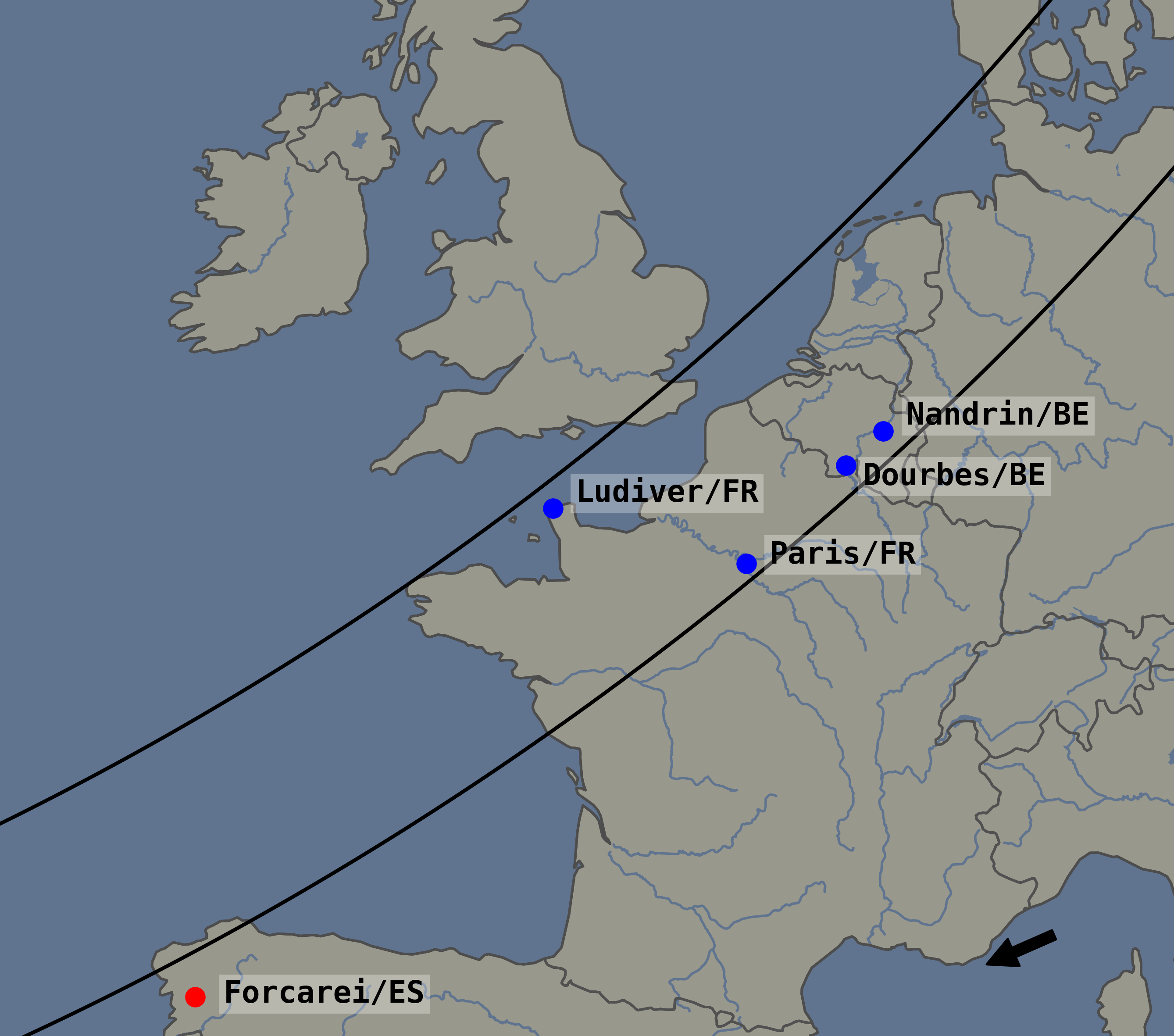}
    \caption{Post-diction map of the event observed on September 8, 2019. The arrow at the bottom right indicates the motion of the shadow. The black lines represent Chiron's shadow path, the blue dots are the sites that detected the occultation, and the red one shows where data were acquired but no event was detected.}
    \label{map2019}
\end{figure}

An observation campaign was organized and observations were eventually made from five sites.  The observation setups can be found in Table \ref{tabobs2019}. Two sites in France (Ludiver and Paris) and two in Belgium (Nandrin and Dourbes) positively detected the event (Fig. \ref{map2019}). Unfortunately, the Belgian sites were in the same line with respect to Chiron's shadow path. Therefore, even if both were used, only three effective chords were available to derive the object's visible limb at the event time. Observations from Forcarei/ES were negative, that is, it did not detect the occultation by Chiron since it was south of the actual shadow path.

\begin{table*}
\caption{Observational circumstances of the September 8, 2019 Chiron occultation.}             
\label{tabobs2019}      
\centering          
\begin{tabular}{l c c c c c c c c }     
\toprule \toprule       
\multirow{2}{*}{Site} & \multirow{2}{*}{Longitude E} & \multirow{2}{*}{Latitude N} & Elevation & Telescope & \multirow{2}{*}{Camera} & Cycle{$^\dag$} & \multirow{2}{*}{$\sigma_{flux}$} & \multirow{2}{*}{Observers} \\
 & & & (m) & aperture (mm) & & time (s) &  
 \\
\midrule        
\multirow{2}{*}{Nandrin/BE}   
& \multirow{2}{*}{05 26 29.5} 
& \multirow{2}{*}{50 31 24.8}    
& \multirow{2}{*}{261}    
& \multirow{2}{*}{406}    
& \multirow{2}{*}{Watec-910HX}     
& \multirow{2}{*}{2.56}   
& \multirow{2}{*}{0.16}  
& O. Schreurs, \\
& & & & & & & & E. Fernandez \\
\multirow{2}{*}{Dourbes/BE} & 
\multirow{2}{*}{04 34 56.0}  & 
\multirow{2}{*}{50 05 25.9}  & 
\multirow{2}{*}{195} & 
\multirow{2}{*}{400} & 
\multirow{2}{*}{Watec-910HX}   & 
\multirow{2}{*}{1.28}  & 
\multirow{2}{*}{0.23} & 
\multirow{2}{*}{R. Boninsegna}  \\
\\ 
\multirow{2}{*}{Ludiver/FR}   
& \multirow{2}{*}{-01 43 42.1} 
& \multirow{2}{*}{49 37 46.9}    
& \multirow{2}{*}{179}    
& \multirow{2}{*}{400}    
& \multirow{2}{*}{RAPTOR Kite}     
& \multirow{2}{*}{0.50}   
& \multirow{2}{*}{0.42}  
& F. Colas, D. Bihel, \\
& & & & & & & & J. Lecacheux  \\
\multirow{2}{*}{Paris/FR}   & 
\multirow{2}{*}{02 20 26.5}  & 
\multirow{2}{*}{48 49 26.9}  & 
\multirow{2}{*}{79}  & 
\multirow{2}{*}{250} & 
\multirow{2}{*}{ASI174MM}    & 
\multirow{2}{*}{1.50}  & 
\multirow{2}{*}{0.33} & 
\multirow{2}{*}{J. L. Dauvergne} \\
\\
\multirow{2}{*}{Forcarei/ES} & 
\multirow{2}{*}{-08 22 15.2} & 
\multirow{2}{*}{42 36 38.4}  & \multirow{2}{*}{670} & 
\multirow{2}{*}{507} & 
\multirow{2}{*}{Watec-910HX}   & 
\multirow{2}{*}{2.56}  & 
\multirow{2}{*}{0.11} &
\multirow{2}{*}{H. González} \\
\\
\bottomrule  
\end{tabular}
\tablefoot{\footnotesize{$^\dag$Read-out time is negligible for these cameras, so the exposure time equals the cycle time.}}
\end{table*}

The occultation light curves (see Appendix \ref{LCs}) were obtained using differential aperture photometry using the Package for the Reduction of Astronomical Images Automatically (PRAIA) \citep{Assafin2011}.  
The ingress and egress times were calculated by modeling the light curve considering a sharp-edge occultation model convolved with Fresnel diffraction, the apparent stellar diameter at the object's distance, and the finite exposure time of each data set \citep[more details can be found in][]{Braga-Ribas2013, Souami2020}. 
The shortest integration time was 0.5 seconds at the Ludiver site, translating into 11.4 km per data point along the track. Considering that the stellar apparent diameter equals 0.08~km, estimated using its B, V, and K magnitudes and \citet{vanbelle1999} equations for a main sequence star, and the Fresnel scale at $D = 17.85$~au (astronomical units) is $L_f = \sqrt{\lambda D / 2} = 0.932~$km, the light curves are smoothed mainly by the exposure time.

The timings of the event are presented in Table~\ref{times2019}. Time-stamping was ensured to be referenced to the Universal Time Coordinate (UTC) on the order of a millisecond by Global Positioning System (GPS) devices for all stations, except at Paris/FR, which had no reliable time source. The observation at Planetarium Ludiver/FR was made using a Raptor CMOS Kite camera and TimeBox system \citep{Gallardo21} and registered the images in ser\footnote{Simple uncompressed video format for astronomical capturing: \url{http://www.grischa-hahn.homepage.t-online.de/astro/ser/}} format. Nandrin/BE site used a GPSBoxSprite-VTI, and Dourbes/BE site used an International Occultation Timing Association Video Time Inserter (IOTA-VTI) system\footnote{\url{https://occultations.org/}}, both with Watec 910HX video cameras, and the avi video format was used for registering the images. Paris/FR used a ZWO ASI174MM camera and registered the event using ser file format. All the data sets were converted to individual fits frames using Tangra software version 3.7.3\footnote{\url{http://www.hristopavlov.net/Tangra3/}}. The video GPS times are inserted at the end of the video camera exposures, so the time of individual frames were subtracted by half an exposure time to give the time of the middle of the integration correctly 
\citep[details in][]{Benedetti-Rossi_2016, Barry2015}.

\begin{table*}[!h]
\caption{Event times and chord length of the four positive sites on September 8, 2019 occultation.}   
\label{times2019}  
\centering          
\begin{tabular}{l c c c }      
\toprule \toprule       
\vspace{1pt}Site & Ingress (UTC) & Egress (UTC) & Chord length (km) \\
\midrule     
\vspace{1pt}Dourbes/BE  & 23:04:12.10  $\pm$ 0.42  & 23:04:17.79  $\pm$ 0.70  & 138 $\pm$ 13  \\
\vspace{1pt}Nandrin/BE    & 23:04:08.00  $\pm$ 0.43  & 23:04:13.40  $\pm$ 0.92  & 146 $\pm$ 28  \\  
\vspace{1pt}Ludiver/FR  & 23:04:27.96  $\pm$ 0.22  & 23:04:33.94  $\pm$ 0.42  & 135 $\pm$ 11   \\
\vspace{1pt}Paris/FR$^\dag$& 23:04:20.39  $\pm$ 0.56  & 23:04:25.47  $\pm$ 0.79  & 116 $\pm$ 22     \\
\bottomrule 
\end{tabular}
~\tablefoot{$^\dag$Times not shifted, see text.}        
\end{table*}

\begin{figure}[!h]
    \centering
    \includegraphics[width=\hsize]{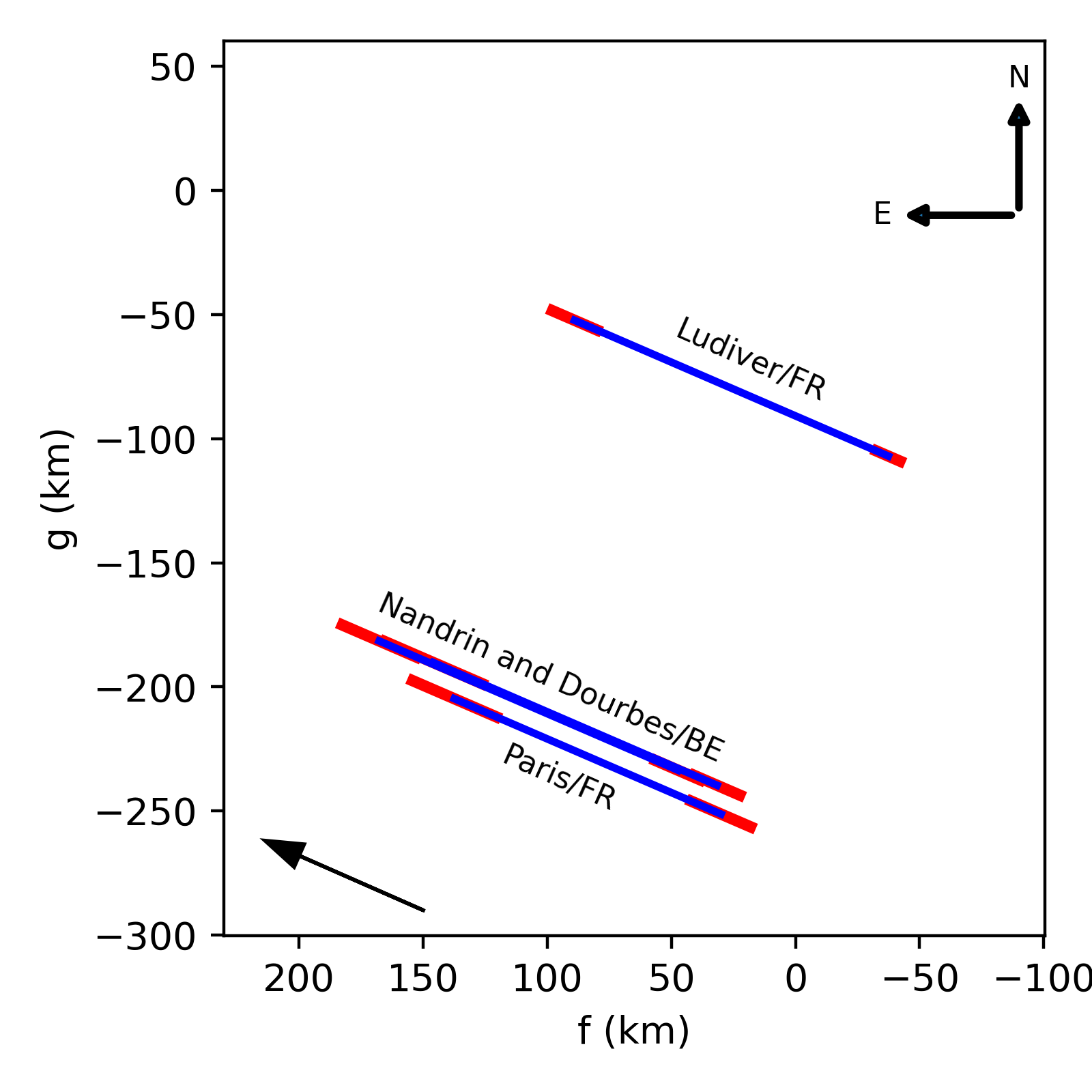}
    \caption{Occultation chords in blue, with error bars in red, observed on September 08, 2019. From north to south, we have the chords from Ludiver/FR, Nandrin, and Dourbes/BE (same line), and Paris/FR. \Gaia DR3 propagated star position and NIMAv10 provide the f{$_0$} and g{$_0$} during the geocenter's closest approach.}
    \label{chords2019}
\end{figure}

\subsection{Fit of the elliptical limb}

The corrected times were projected in the sky plane and used to determine Chiron's limb shape at the time of the occultation. It is possible to see in Fig. \ref{chords2019} that the Paris chord is clearly shifted with respect to the other sites. As it did not benefit from a reliable time source, we shifted it forward by 0.9 seconds by minimizing the dispersion of the fitted ellipses with respect to the chord extremities. Due to the distribution and errors of the detected chords, a wide range of ellipses can be fitted. 

The most recent and accurate work on determining Chiron's size, available in the literature, combined thermal fluxes with mid/far infrared fluxes to derive Chiron's relative emissivity at radio (mm and sub-mm) wavelengths \citep{Lellouch2017}. We considered their spherical approach, which gives an area-equivalent diameter of 210$^{+11}_{-14}$ km, as we understand it to be the most conservative one. For instance, the error bar of the size given by this model encloses all the other scenarios (e.g., different roughness and elliptical profile). We used the aforementioned equivalent diameter as upper and lower limits on our fitting procedure to restrict the range of possible elliptical limbs compatible with our observed chords.

Our elliptical limb shape fitting to the occultation chords is detailed in \citet{Braga-Ribas2013} and \citet{Souami2020}, for example.
In short, we generated a large set of ellipses, varying 
their semi-major axes $a'$ from 90 to 160 km by steps of 1 km, 
their oblatenesses $\epsilon= (a'-b')/a$ (where $a'$ and $b'$ are the semi-major and semi-minor axis of the apparent ellipse) 
from 0 to 0.625 by steps of 0.005, and 
their Position Angles (PA)\footnote{The Position Angle is the angle between the celestial north and the minor-axis orientation, measured positively to the east.} from -90$^\circ$ to +90$^\circ$ by steps of 2$^\circ$.
Moreover, in this procedure, we imposed that the area-equivalent radius, 
$R_{\rm equiv}=a\sqrt{1-\epsilon}$, of the ellipse be consistent with the 
\citet{Lellouch2017}'s value within the error domain mentioned above, that is, 
$98 \leq R_{\rm equiv} \leq 111$~km. 

\begin{figure}[!h]
    \centering
   \hspace{-0.5 cm}
    \includegraphics[width=\hsize]{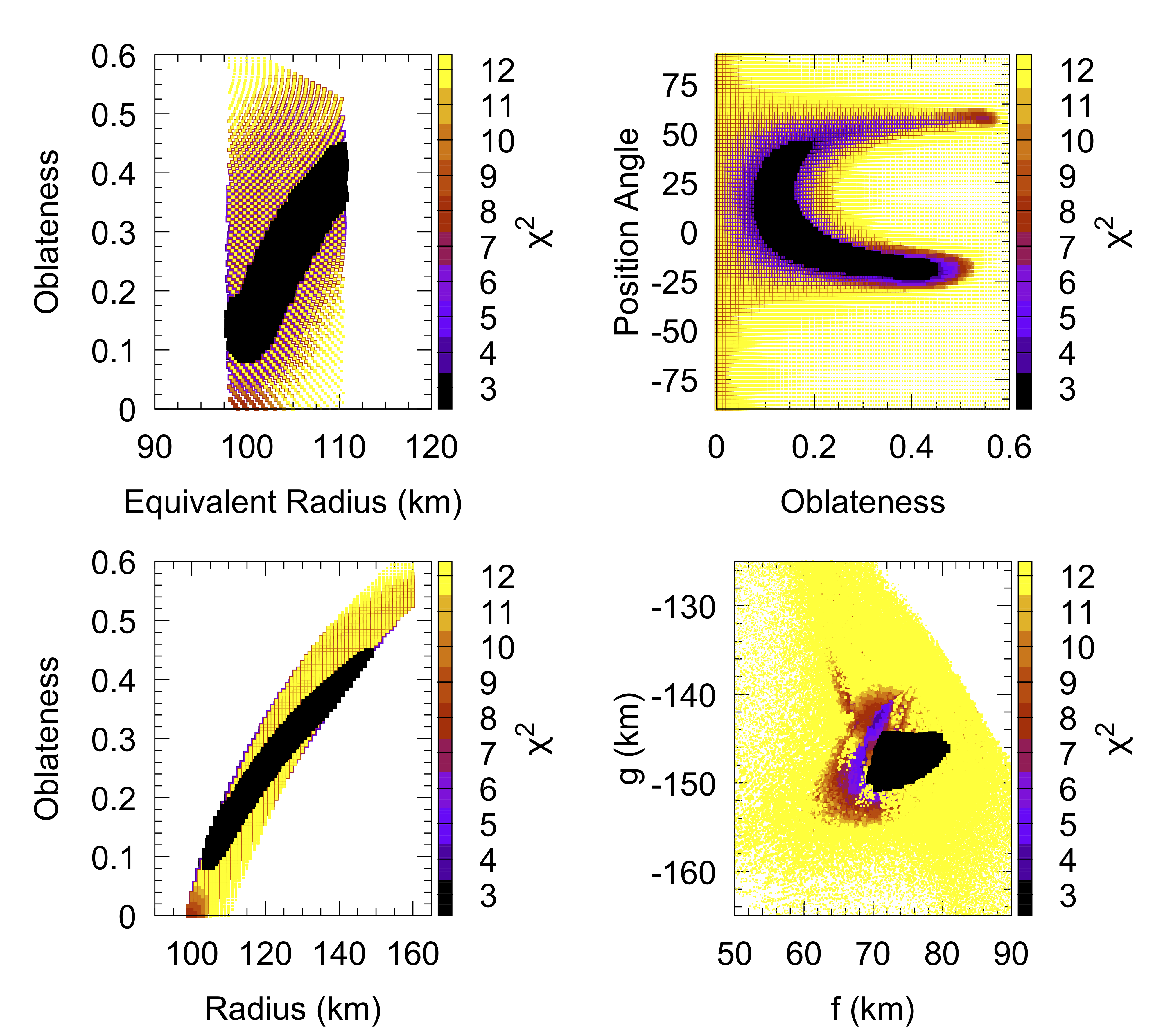}
    \caption{ Combinations of the limb-fitting parameters shown in each panel.
    The 1$\sigma$-level solutions are plotted in black. The solutions above this limit are shown in different colors, as defined in the color bar. Solutions above the 3-$sigma$ are shown in yellow.}
    \label{chi2map}
\end{figure}

For each ellipse, the free parameters of the fit were the coordinates of
body center $(f\rm_c,g\rm_c)$ projected in the sky plane.
This fit returned the classical $\chi^2$ value based on the radial residuals of the fit, that is, the radial distance of the chord at the ingress/egress, $r_{i,obs}$, to the tested ellipse, $r_{i,cal}$, proportional to the radial uncertainty, $\sigma_{radial}$, of the ingress/egress time (Eq. \ref{chi2}).
The results were sorted from the minimum value $\chi^2_{\rm min}$, and for all the ellipses up to $\chi^2+1$ to define the marginal 1$\sigma$ error bar of each parameter.
The results are displayed in Fig.~\ref{chi2map} for various parameters of interest.


\begin{equation}
 \label{chi2}
\chi^2 = \sum_{\rm i=1}^{\rm N} \frac{({r_{i,obs}} -{r_{i,cal}})^2}{{\sigma_{radial}}^2}
.\end{equation}

\begin{figure}[!h]
    \centering
    \includegraphics[width=\hsize]{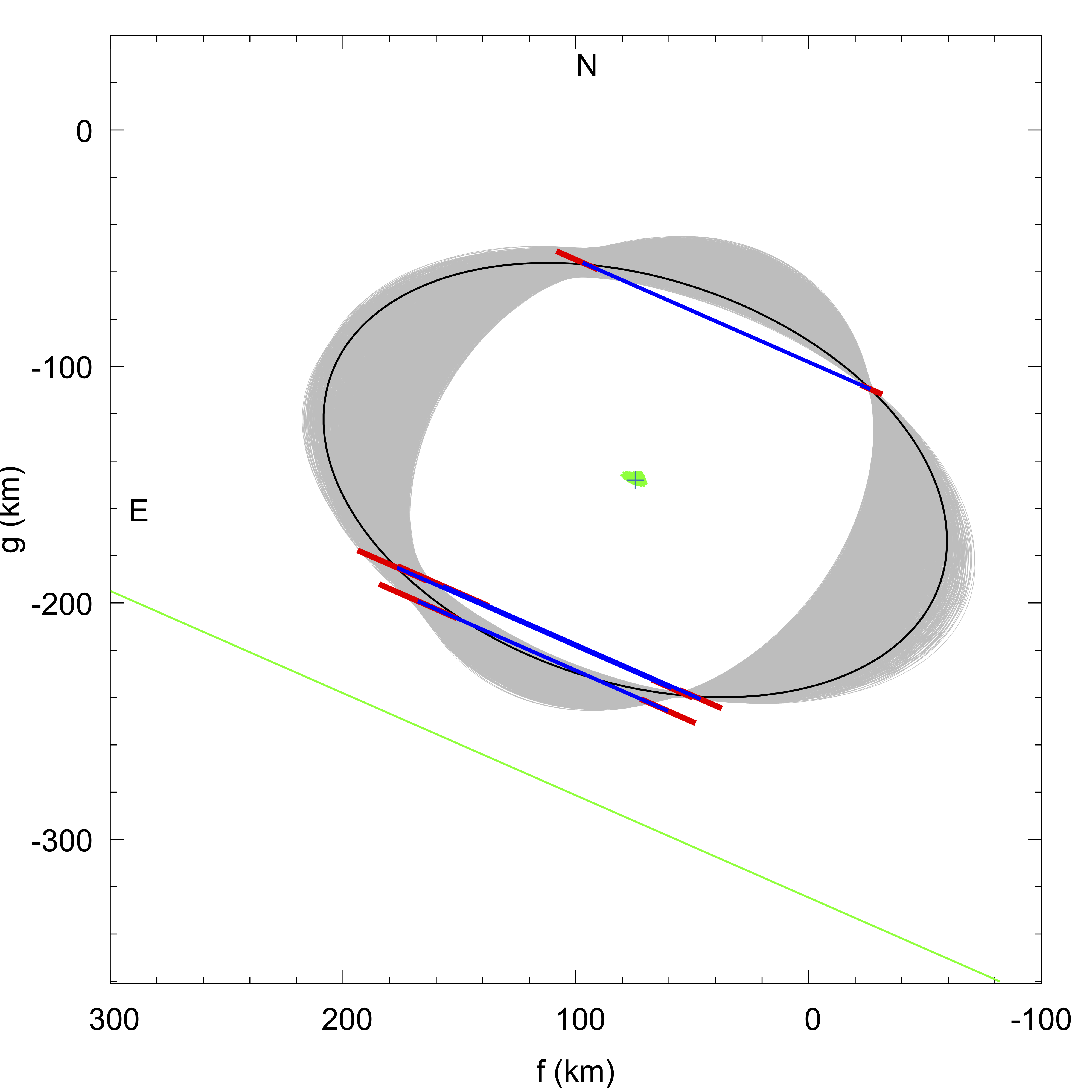}
    \caption{Same as Fig.~\ref{chords2019}, with the best fitting ellipse plotted in black (e.g., the lowest ${\chi^2}$ value) and the 1$\sigma$-level ellipse plotted in grey. The green spot in the center of the ellipses represents their respective centers. The Paris chord was shifted forward by 0.9 seconds for the best alignment with the other sites (see text for details). The green line below the ellipses is the negative chord observed from the Forcarei site.}
    \label{ellipses}
\end{figure}

Table~\ref{result2019} summarizes the results of the fit (limb's size, shape, and orientation) and 
Fig.~\ref{ellipses} displays the best ellipse in black and the 1$\sigma$-level solutions in gray.

\begin{table}[h]
    \centering
    \caption{Elliptic fit to the Chiron's limb from the September 8, 2019 occultation.}
    \begin{tabular}{l l} \toprule \toprule  
    Equivalent radius$^\dag$ (km)   & $R_{\rm equiv}= 105^{+6}_{-7}$ \\
    Semi-major axis (km)            & $a= 126 \pm 22$ \\
    Position angle (deg)            & $PA= 11 \pm 33$ \\
    Oblateness                      & $\epsilon$= 0.27 $\pm$ 0.18 \\ 
    \bottomrule
    \end{tabular}
    \vspace{-0.2cm}
    \tablefoot{$~^\dag$\small{Taken from \citet{Lellouch2017}, see text.}}
    \label{result2019}
\end{table}

\subsection{Size and shape constraints}

Although challenging due to the presence of coma and ring, rotational
light curves have been observed for Chiron, giving a
synodic rotation period of P = 5.917813~$\pm$~0.000007 h \citep{Marcialis1993}. \citet{Groussin2004} derived a true light curve amplitude
of $\Delta m_0=~$0.16~$\pm$~0.03, which is the light curve amplitude that we would observe if the object's pole is perpendicular to the line of sight. This implies an axial ratio 
$b/a=\beta= 0.862\pm0.023$, for a triaxial body with axes $a>b>c$. Considering the pole direction
given by \citet{Ortiz2015} to be the same as that of the
main body, the opening angle (B) with respect to the object’s equator
was B = 42$^\circ$ in 2019. So, the observed semi-major axis $a$ of the ellipse can be considered
as the actual equatorial radius of Chiron within a factor of 1.16 (at maximum), given the known rotational variability
and the fact that the rotational phase of Chiron at the time of the occultation was not known, implying that $b = 109\pm19~$km.

If Chiron has a Jacobi shape (i.e., fluid hydrostatic equilibrium), we can expect $c/a=\gamma = 0.53905\pm0.00785$ for a plausible density of  $1,119~\pm~4~\rm{kg~m^{-3}}$  and rotation period of P = 5.917813~$\pm$~0.000007~h, which would give a semi-minor axis of $c=$~68~$\pm$~13~km, using \citet{Chandrasekhar1987} formalism.
Under these assumptions, Chiron has a volume-equivalent radius of R$_{vol}$= 98~$\pm$~17~km.

\begin{figure}[!h]
   \centering
\includegraphics[width=\hsize]{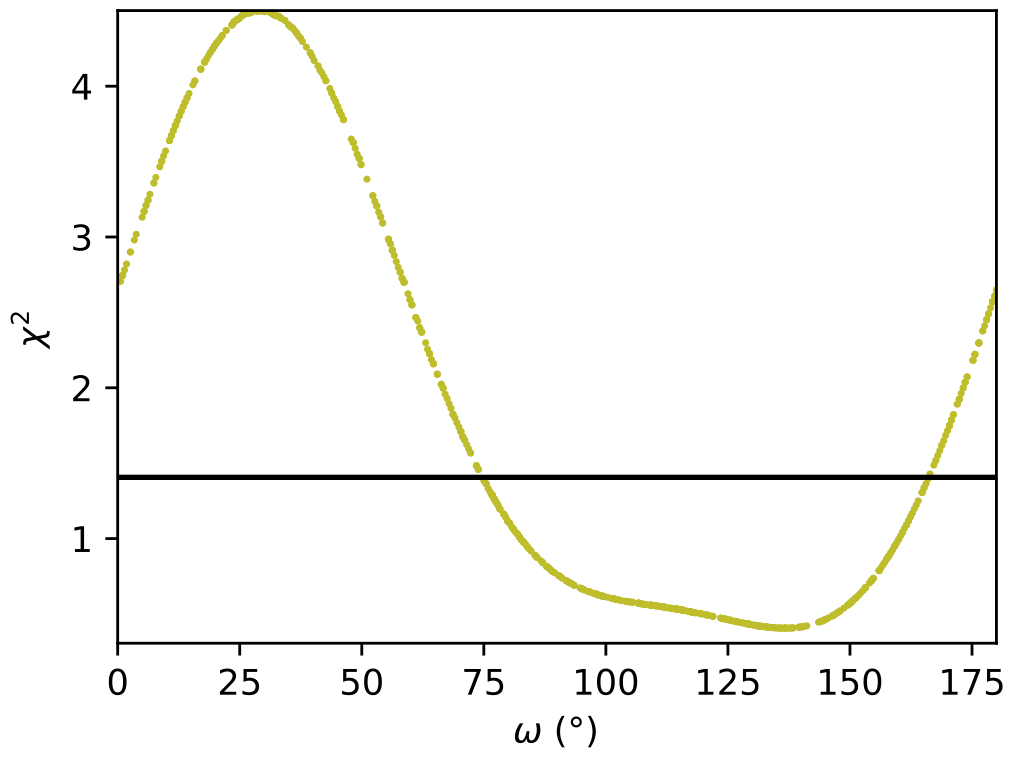}
    \vspace{-0.2cm}
    \caption{Test of all possible rotation phases, considering Chiron as a Jacobi ellipsoid. Rotation phases within a range of 90$^\circ$ are compatible.}
    \label{w0}
\end{figure}

\begin{figure}[!h]
\vspace{-0.4cm}
   \centering
\includegraphics[width=\hsize]{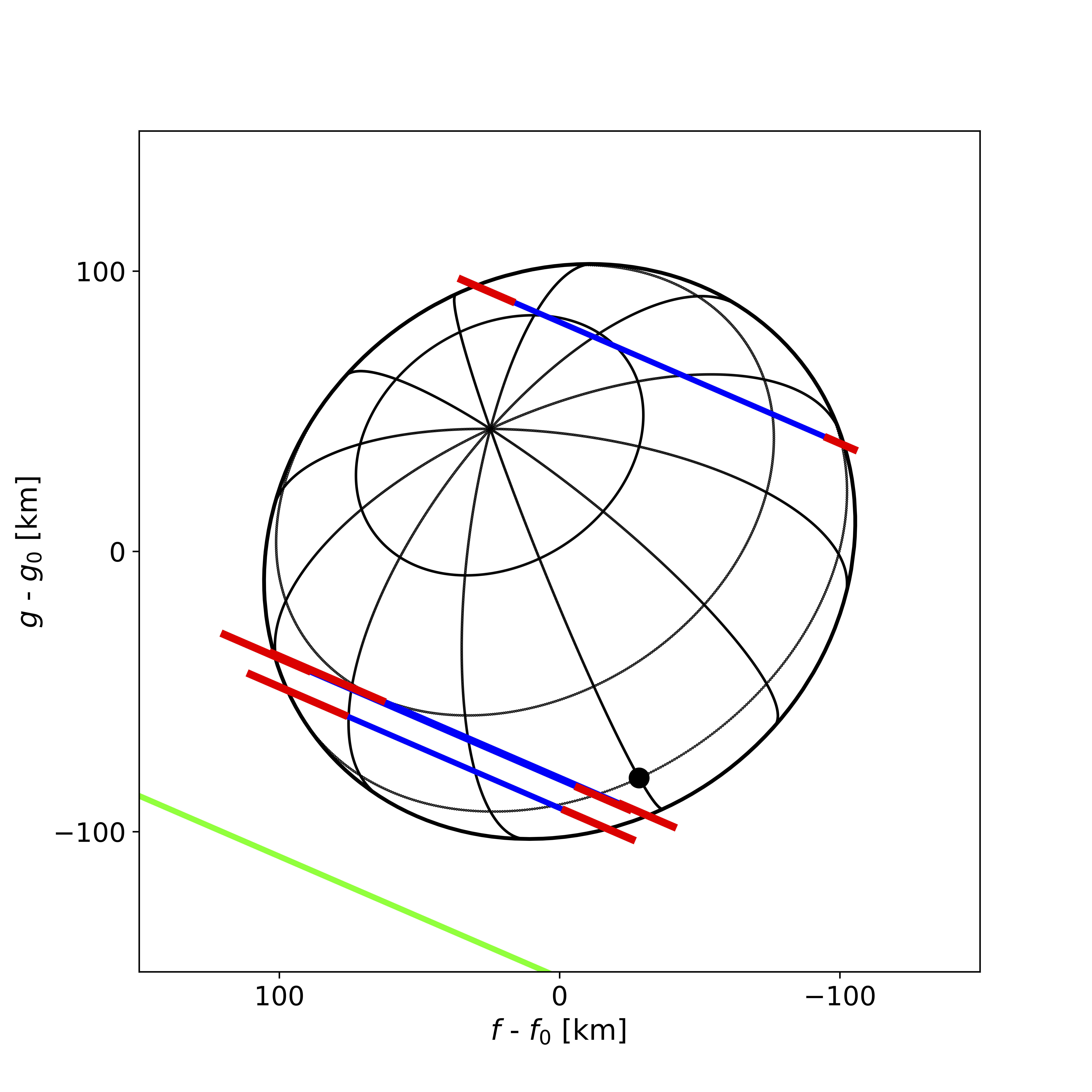}
    \vspace{-0.7cm}
    \caption{Best fit of the Jacobi ellipsoid with $a=126\pm22$~km, $b=109\pm19$~km, and $c=68\pm13$~km derived for Chiron, with a $\omega=120^\circ$, represented by the big black dot, which also indicates the position of the major-axis $a$. Here, f{$_0$} and g{$_0$} are the center of the ellipse presented in Fig.\ref{ellipses}.}
    \label{jacobi}
\end{figure}

We verified if the proposed ellipsoid agrees with the observed chords by assuming that Chiron has the same pole as the rings proposed by \citet{Ortiz2015} (i.e., the ring's plane crosses the object's equator). As the rotational phase during the 2019 event is unknown, we must test all the phases. 
Being $\omega$ the angle of the object's major-axis ($a$) with respect to the intersection of the object's equatorial plane and the International Celestial Reference System (ICRS) fundamental plane, as defined in \citet{Archinal2018}, we note that values of $\omega$ in a range of 90$^\circ$ (i.e., one forth of a phase) provides compatible fits with the observed chords (Fig. \ref{w0}). Figure \ref{jacobi} shows the best fit of Chiron as a Jacobi ellipsoid fitted to the 2019 occultation chords.


\section{Stellar occultation of November 28, 2018}
\label{2018occ}

Also predicted by the Lucky Star project, this event was of interest due to the small shadow velocity of only 5.77 $\rm{km~s^{-1}}$, relative to the geocenter, thus allowing a high spatial resolution per data point. The occulted star was the \Gaia DR3 2646598228351156352, with a magnitude G = 17.28. Propagating the position of the star from the \Gaia DR3 catalog epoch using parallax and proper motion to the time of the event, its geocentric position at epoch is:

\vspace{-5mm}
\begin{center}
$$ \alpha = 23h~46m~04s.33575 \pm 0.2508~{\rm mas,} $$ 
\vspace{-5mm}
$$ \delta = +02^\circ~13'~05''.5270 \pm 0.3445~{\rm mas.} $$
\end{center}

\begin{figure}[!h]
    \centering
    \includegraphics[width=\hsize]{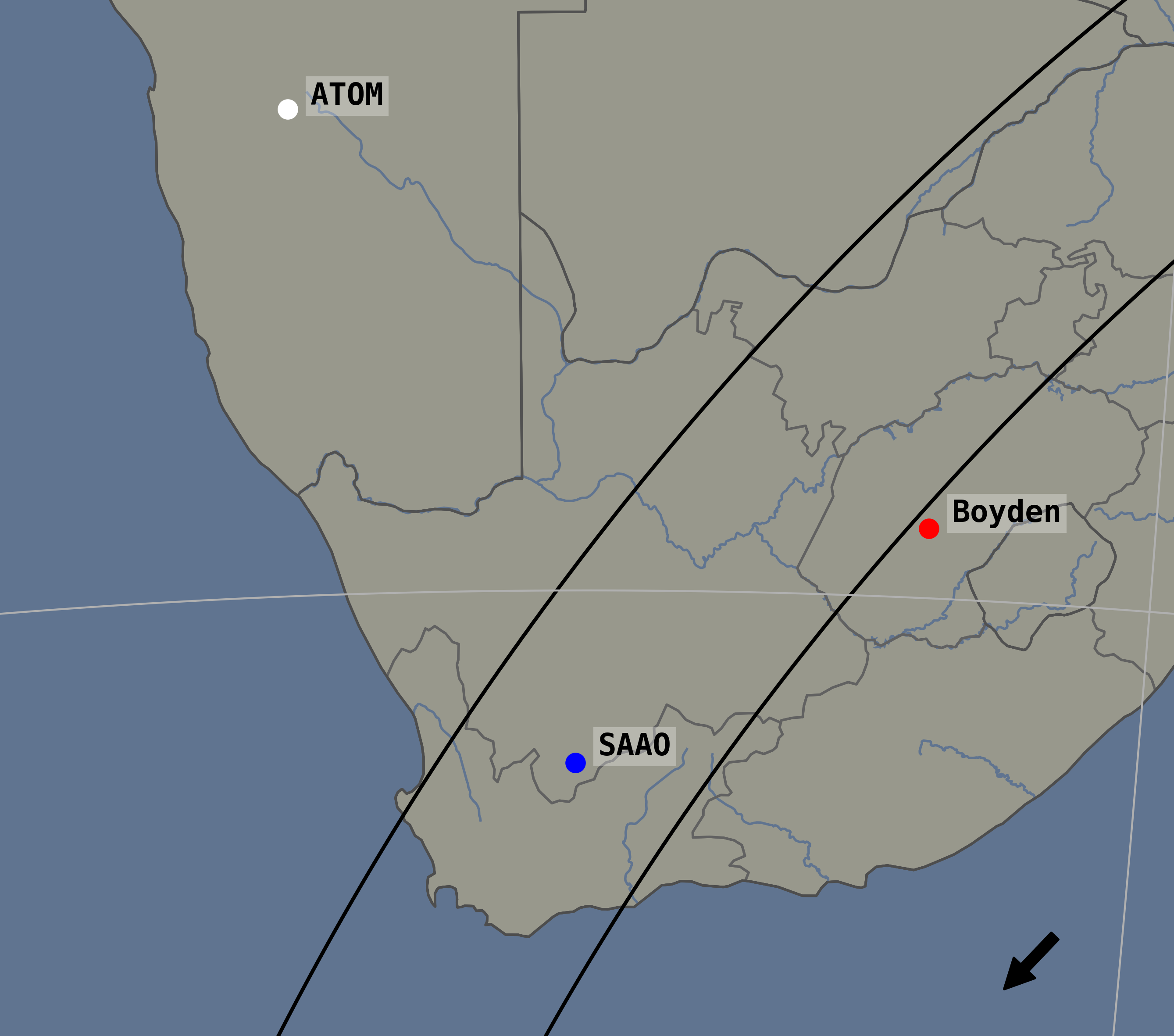}
    \caption{Post-diction map of the November 28, 2018 event. 
    The black lines delineate Chiron's shadow path, 
    the blue dot is the SAAO site where the occultation was detected, 
    the red dot is Boyden, where data were acquired but no event was detected, and 
    the white dot (ATOM) could only start acquisition after the actual local closest approach.}
    \label{map2018}
\end{figure}

Observations were arranged at 
the SAAO and 
at the Boyden Observatory, both in the Republic of South Africa.
A third site was used in Namibia, the Automatic Telescope for Optical Monitoring (ATOM); {\cite[See][for details on this facility.]{Hauser2004}}. 
Table~\ref{tabobs2018} provides circumstantial details and 
Fig.~\ref{map2018} displays the post-diction map. 
 
\begin{table*}
\caption{Observational circumstances of the November 28, 2018 Chiron occultation.}             
\label{tabobs2018}      
\centering          
\begin{tabular}{l c c c c c c c c }     
\toprule \toprule      
\multirow{2}{*}{Site} & Longitude E & Latitude S & Elevation & Telescope & \multirow{2}{*}{Camera} & Cycle$^*$ & \multirow{2}{*}{$\sigma_{flux}$} & \multirow{2}{*}{Observers} \\
 & ($^\circ$ ' '') & ($^\circ$ ' '') & (m) & aperture (mm) & &  time (s) &  \\
\midrule            
SAAO    & 20 48 37.8  & 32 22 41.9  & 1,760 & 1,016 & SHOC$^\ddagger$         & 1.5  & 0.08 & A. Sickafoose \\
Boyden  & 26 24 17.0  & 29 02 19.8  & 1,372 & 1,500  & Alta-U55    & 10   & 0.08 & P. van Heerden  \\
ATOM$~^\dag$    & 16 30 05.6  & 23 16 19.7  & 1,800 & 700  & DU888\_BV    & 1.0   & 0.20 & F. Jankowsky  \\
\bottomrule                 
\end{tabular}
\tablefoot{\footnotesize{$^*$Read-out times are negligible for these cameras, so the exposure time equals the cycle time. $^\dag$Observation initiated after the actual local closest approach. $^\ddagger$SHOC stands for Sutherland High-speed Optical Cameras, see \citet{Coppejans2013}. }
}
\end{table*}

\begin{table}[h]
\caption{Derived event times and chord length as observed from SAAO on November 28, 2018.}             
\label{times2018}  
\centering          
\begin{tabular}{l  l }      
\toprule \toprule    
Site & SAAO/SA \\
Ingress (UTC) & 20:50:12.51 $\pm$ 0.06 sec \\
Egress (UTC)  & 20:50:43.80  $\pm$ 0.10 sec\\
Chord length  & 180.5 $\pm$ 0.9 km \\
\bottomrule    
\end{tabular}
\end{table}

The occultation could only be detected from SAAO.
Using an exposure time of 1.5 seconds and no filter, each data point has a spatial resolution of 8.65~km and S/R=14. Using the same procedures described in Section \ref{2019occ}, the occultation light curve and the ingress and egress times were obtained (see Table \ref{times2018}). The stellar apparent diameter was estimated to be 0.3~km
and the Fresnel scale at $D= 18.3538$~au is $L_f = 0.945$~km, so, again, the light curve smoothing was largely dominated by the exposure time. 

The chords in the sky plane are displayed in Fig. \ref{rings2018}. The large exposure time (10 seconds) used at Boyden Observatory smooths any putative short event, making it undetectable. Considering the long chord observed from SAAO, a long chord (at least one exposure time) would be detected from Boyden as well. As no flux drop was seen from there (Fig. \ref{fig:LC}), we can conclude the center of Chiron was north of SAAO. This is a crucial constraint for the search of secondary events caused by the putative rings proposed by \citet{Ortiz2015}.

\section{Constraints on the secondary events}

As mentioned earlier in this work, secondary events have been observed in the past during stellar occultations by Chiron \citep{bus1996, elliot1995, Ruprecht2015} and it has been suggested that they may have been caused by a ring system, jets, or a dust shell \citep{Ortiz2015, Sickafoose2020}. The only data set presented in this paper that offers useful limits on the detection of material around Chiron, while considering the image cadence and S/N, is the one obtained at SAAO in November 2018 (discussed further in this section).

\subsection{Reanalysis of 2011 data} \label{sec:rings2011}

We re-assessed the physical properties of the features observed during the occultation by Chiron in 2011 \citep{Ruprecht2015} when symmetrical secondary drops were detected around Chiron. Under the assumption that they are rings, as proposed by \citet{Ortiz2015}, we followed the same procedures as those we used to analyze (10199) Chariklo's rings; \cite[see][]{Braga-Ribas2014, berard2017, Morgado2021}.

\begin{figure}[!h]
    \centering
    \includegraphics[width=\hsize]{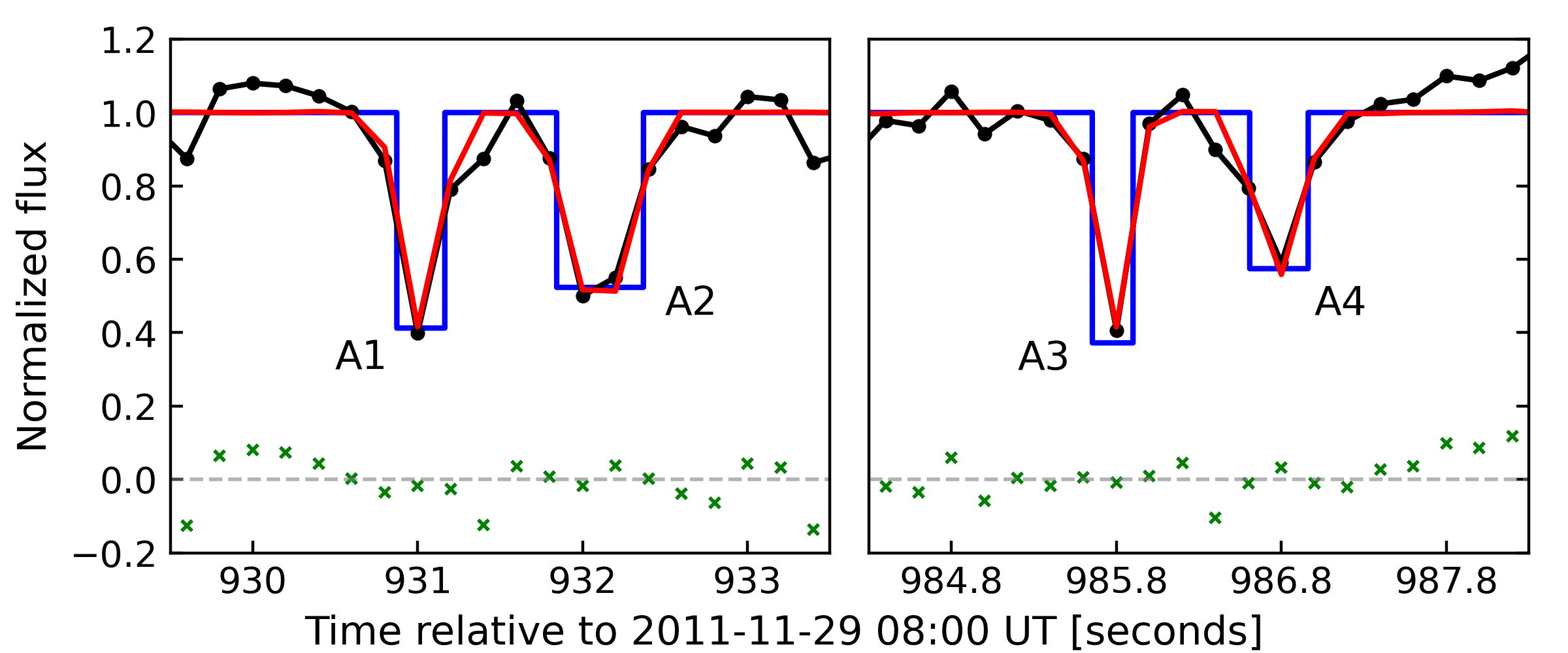}
    \caption{Reanalysis of the FTN data obtained during the November 29, 2011 stellar occultation by Chiron; \citep{Ruprecht2015, Sickafoose2020}.
    Adopting the preferred pole position proposed on \citet{Ortiz2015}, we fit the light curves with two sharp-edged semi-transparent rings, accounting for the stellar velocity perpendicular to the ring as measured in the ring plane. Data are shown in black, the ring models are plotted in blue, and the red curves are the synthetic light curves, accounting for diffraction, stellar diameter, and finite integration time. The green crosses at the bottom are the fit residuals to the synthetic light curve. The times were corrected by -7 seconds with respect to the values published in \citet{Ruprecht2015} and \citet{Sickafoose2020}, so they are back to their original values (see text).}
    \label{ringfit2011}
\end{figure}

Using the FTN data, as presented in Fig.~4 of \citet{Ruprecht2015} and reproduced in Fig.~2 of \citet{Sickafoose2020}, we first corrected the timing of those plots to their original values, which are presented in the above-mentioned references with a shift forward of 7 seconds to account for the geographic offset between FTN at Haleakala and IRTF at Mauna Kea. 
Using the preferred pole position of \citet{Ortiz2015}, with ecliptic coordinates $\lambda=144^o,~\beta=24^o$, we calculated the opening angle with respect to the object’s equator (i.e., the sub-observer point),
which was B = 59.6$^\circ$ in 2011, and the relative star velocities perpendicular to the rings, as measured in the ring plane.  With these values, we fitted two sharp-edged rings with different sizes and opacity, accounting for diffraction \cite[see the Extended Data in][for further details]{Braga-Ribas2014,berard2017}.
We tested ring widths varying from 0.5 to 4 km and opacity from 0 to 0.6.
The resulting fits are shown in Fig.~\ref{ringfit2011}.

\begin{table*}[h!]
\centering
\small
\caption{Ring properties obtained from Faulkes 2011 data. We used the same nomenclature as used by \citet{Sickafoose2020}, where A1 and A2 correspond to the drops before the closest approach and A3 and A4 after the closest approach.}
\label{tab:2011rings}      
\begin{tabular}{l c c c c} 
\toprule \toprule
\multicolumn{5}{c}{\bf{Ring properties from the 2011 detection}} \\ 
\midrule
& A1       & A2       & A3       & A4       \\  
\midrule
$v_{\perp ring}$ (km s$^{-1}$) & 7.18 & 6.99 &  8.67 & 8.73   \\
$t_{mid}$ (sec)$^\dag$  & 931.01 $\pm$ 0.02     & 932.06 $\pm$ 0.02  & 985.78 $\pm$ 0.01    & 986.80 $\pm$ 0.02   \\
W$'$ (km)  & 1.53 $\pm$ 0.15  &   2.38 $\pm$   0.20  &   2.07 $\pm$   0.37 &   3.27 $\pm$   0.39 \\
W$\rm_r$ (km) &   2.22 $\pm$   0.21 &   3.63 $\pm$   0.29 &   1.98 $\pm$   0.36  &   3.12 $\pm$   0.37 \\
$\rm{p}'$ &   0.61 $\pm$   0.06  &   0.47 $\pm$   0.04 &   0.62 $\pm$   0.06  &   0.40 $\pm$   0.06 \\
p$\rm_N$  &   0.32 $\pm$   0.04  &   0.23 $\pm$   0.03 &   0.33 $\pm$   0.04  &   0.20 $\pm$   0.03 \\
$\tau_{\rm{N}}$  &   0.41 $\pm$   0.14  &   0.27 $\pm$   0.08  &   0.42 $\pm$   0.16  &   0.22 $\pm$   0.09 \\
E$\rm_p$ (km)  &   0.71 $\pm$   0.11  &   0.85 $\pm$   0.12    &   0.65 $\pm$   0.14 &   0.61 $\pm$   0.12 \\
\bottomrule
\end{tabular}
\vspace{-0.2cm}
~\tablefoot{$^\dag$ Seconds after 08:00 UT. Corrected by -7 seconds; see text. 
} 
\end{table*}

Table 6 shows the perpendicular velocity relative to the rings, $v_{\perp ring}$, the mid-time of each detection, ($t\rm_{mid}$),
the apparent radial width, ($\rm{W}'$), where "apparent" refers to the quantity measured in the sky plane,
the radial width, (W$\rm_r$), in the ring plane,
the apparent opacity, ($\rm{p}'= 1-I/I_0 $),
the normal opacity to the ring plane, (p$\rm_N=|\sin{B}|(1-\sqrt{1-\rm{p}'})$), for a monolayer ring,
the normal optical depth to the ring plane, ($\tau\rm_N=|\sin{B}|(\tau'/2),$ for a poly-layer ring,
where $\tau'=-\ln(1-\rm{p}')$,
and the equivalent width, ($E\rm_p= W\rm_r~\rm{p_N}$). 

They are consistent with the values of \citet{Sickafoose2020} but differ due to i) the applied velocity to calculate the widths and ii) a factor of two between the apparent and the actual ring optical depths
due to light diffraction by individual ring particles, as per \citet{cuzzi1985} and the
discussion in
\citet{berard2017}.
This factor of two applies in the cases where the Airy diffraction scale on individual particles with
radius $r$ $L\rm_A~\sim~(\lambda/2r)D$, is larger than the width of the ring. 
In visible wavelengths and assuming ring particles with sizes of ten meters at most, 
this provides $L \rm_A \gtrsim 80$~km for $D =18.3538$~au, hence validating our assumption (i.e., the actual ring optical depth is half of the one measured with the occultation light curve).
Consequently, our equivalent width  ($E \rm _p$) values are almost half of those given by \citet{Sickafoose2020}.

\subsection{Searching for rings}

\begin{figure}[!h]
    \centering
\includegraphics[width=\hsize]{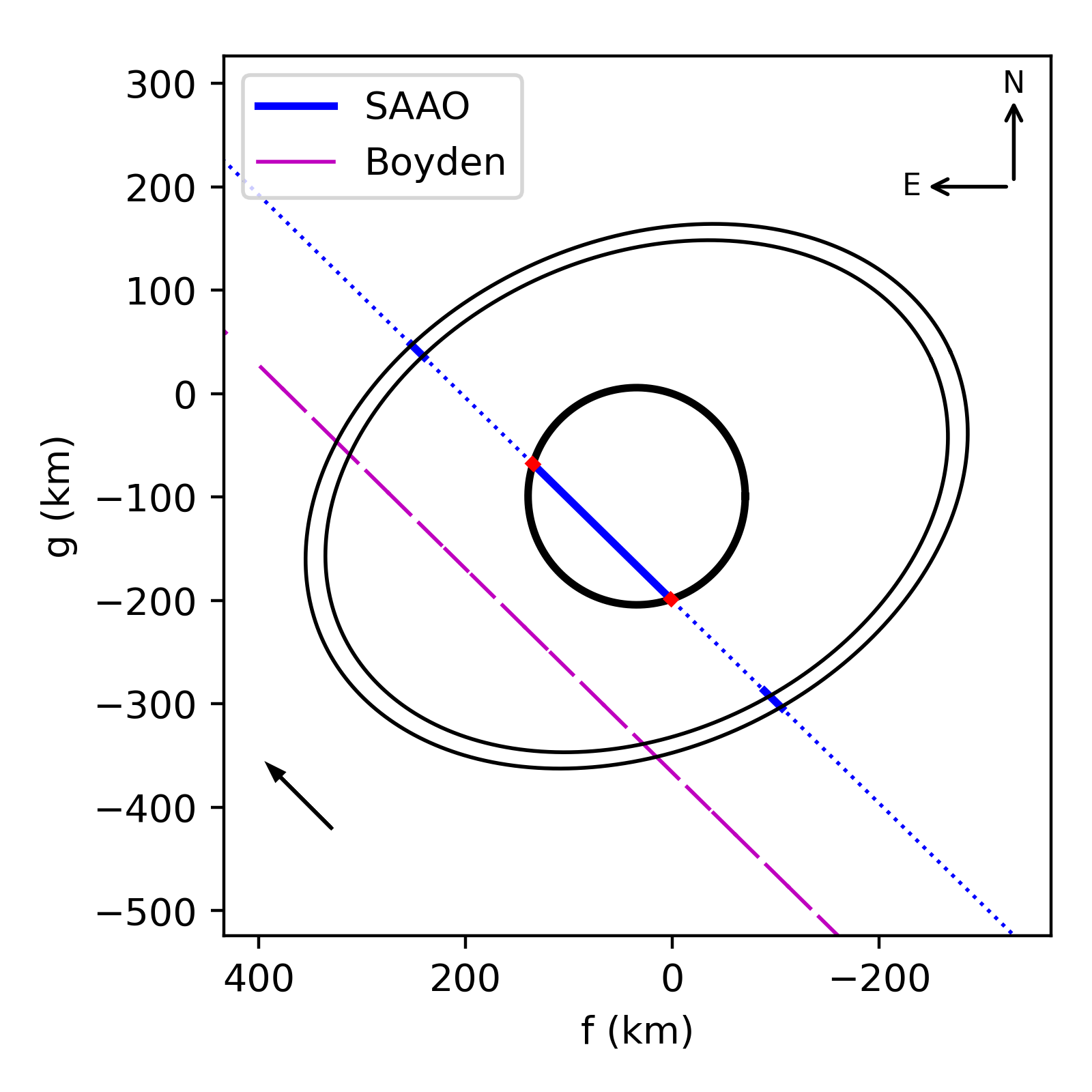}
    \caption{Reconstructed geometry of 2018 event with the rings as proposed by \citet{Ortiz2015}. The large blue segment corresponds to the chord detected from SAAO, and the small ones are the expected location of the rings. The magenta dashed line corresponds to each exposure from the Boyden observatory negative chord.}
    \label{rings2018}
\end{figure}

To search for the proposed rings in the November 28, 2018 SAAO data set, we used the preferred pole position of \citet{Ortiz2015} ($\lambda \sim 144^\circ,~\beta \sim 24^\circ $) and the ring parameters derived from our re-analysis of the 2011 data (Table~\ref{tab:2011rings}). 
From the event geometry (Fig. \ref{rings2018}), we can estimate the time intervals where the proposed rings are expected (small blue segments).

\begin{figure}[!h]
 \includegraphics[width=\hsize]{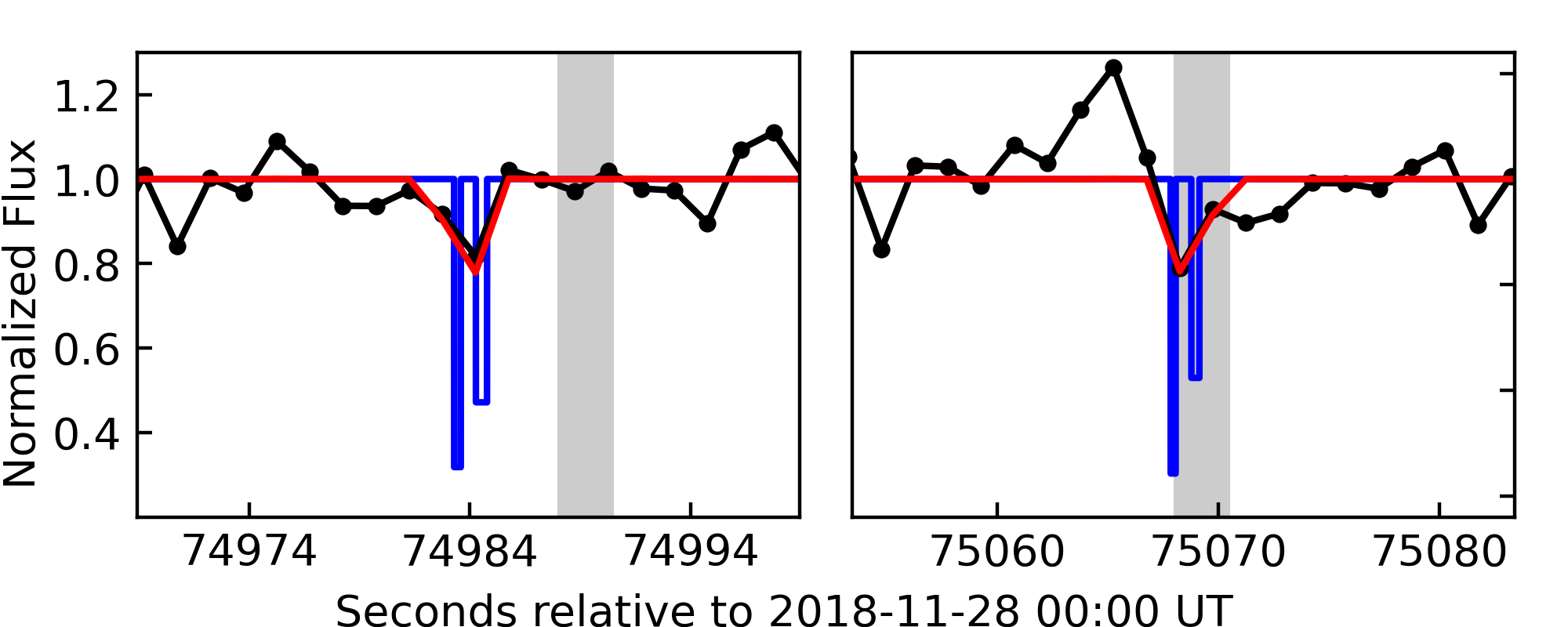}
    \caption{Search for the Chiron's rings proposed by \citet{Ortiz2015} with the properties given by \citet{Sickafoose2020} using the SAAO data obtained during the November 28$^{\rm th}$ 2018 occultation. The gray regions indicate the ring's expected positions from \citet{Ortiz2015}.
    The left (resp. right) panel presents the data before (resp. after) the occultation by the main body.}
    \label{ringsfit2018}
\end{figure}

Using the properties of the rings found in Section \ref{sec:rings2011}, we searched a region of 20 seconds (or $\sim$115 km) around the expected rings locations, on both sides (before and after the main body ingress and egress). This corresponds to about seven times the total duration of the occultations by the rings and gap. The best fit for the ingress is obtained a few kilometers farther than the expected location ($\sim$36 km) and has a $\chi^2$ value per degree of freedom of $\chi^2_{\rm pdf}= 0.59$ (Fig. \ref{ringsfit2018}, left panel). For the egress, the best fits occur at the expected location, with a $\chi^2_{\rm pdf}= 1.09$  (Fig. \ref{ringsfit2018}, right panel). 

\begin{figure}[!h]
    \centering
    \includegraphics[width=\hsize]{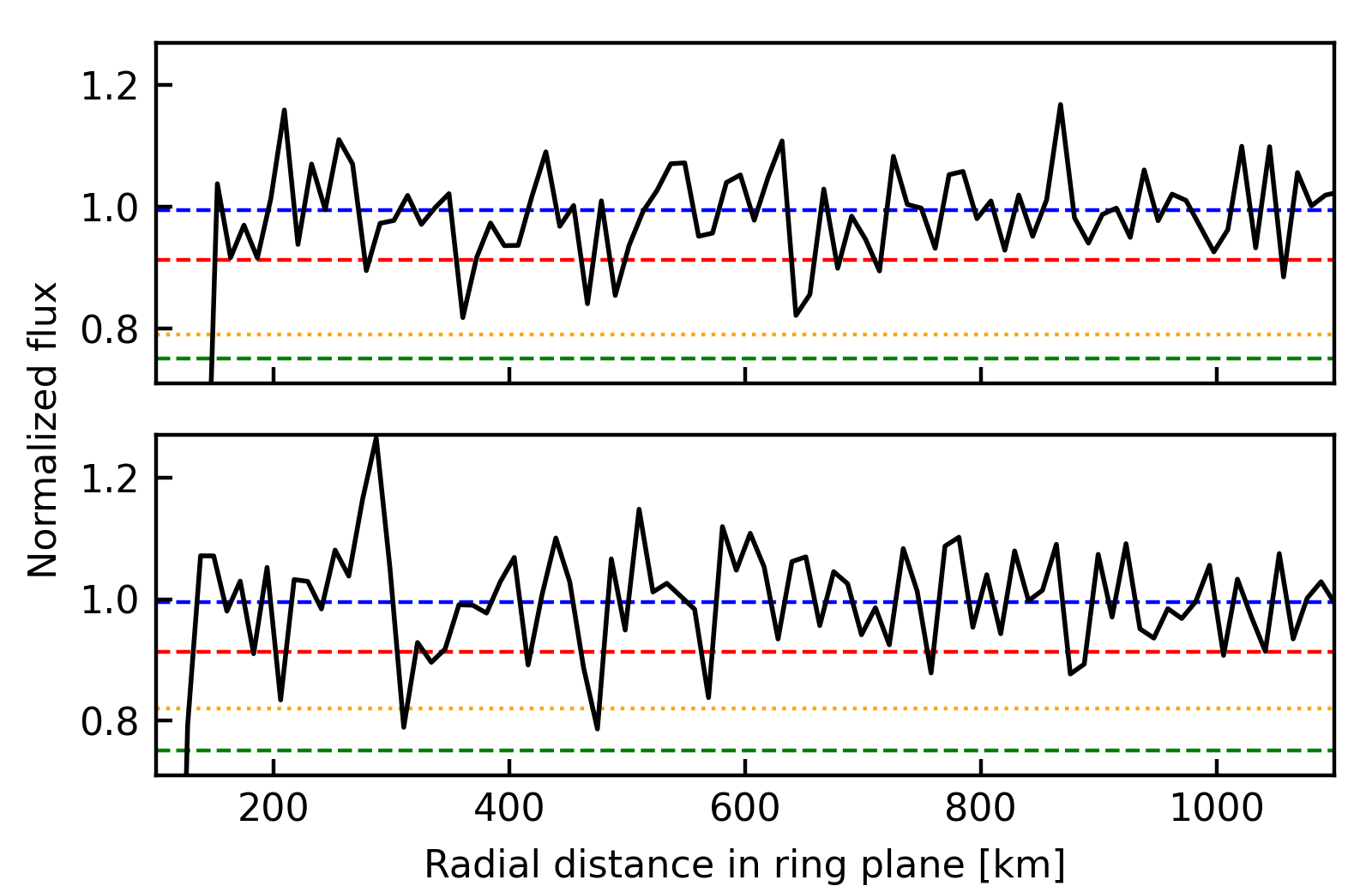}
    \caption{Light curve of the flux versus the distance to Chiron's center in the ring plane. The dashed lines in red and green correspond to the detection limits at the 1 and 3-$\sigma$ levels, respectively. The yellow dotted line corresponds to the expected depth of the occultation promoted by the proposed rings.} 
    \label{opacity2018}
\end{figure}

To assess the detection level of ring material, we analyzed the SAAO light curve, following the same procedures as in \citet{berard2017}. 
We considered the data up to 2,000 km from Chiron's center, excluding the main body occultation part. 
The light curve was converted into flux as a function of radial distance to the object's center, 
counted in the ring plane.
Accounting for the standard deviation of the light curve, 
we obtained a 1$\sigma$ upper limit of $\rm{p}'= 0.08$ for the apparent opacity ($\rm{p}'= 0.24$ at the 3$\sigma$ level), see the red (resp. green) dashed line in Fig. \ref{opacity2018}. 
From the relation $\tau'=-\ln (1-\rm{p}'),$ we derive an upper limit of  $\tau'=~0.25$ (3$\sigma$ level) for the apparent optical depth.

The star velocity at the ring plane was 7.99 $\rm{km~s^{-1}}$, yielding a spatial resolution of 12 km per data point. 
If existent, individual rings of this width would be detectable if $\rm{p}_{\rm{N}} = 0.09$ or $\tau \rm_N > 0.09$. The proposed rings would then be detected at a 2.6$\sigma$ level during ingress and at a 2.2$\sigma$ during egress (see the yellow dashed lines in Fig. \ref{opacity2018}).
This means that the expected ring signatures were too weak to be assertively detectable in this data set with a 3$\sigma$ or higher confidence level.

\subsection{Searching for a shell of material}

The SAAO data can also be used to put upper limits on the detection of a shell of material around Chiron. Using the same procedure as the previous section, the light curve was converted to flux versus distance to the object's center, but here the data were projected to the sky plane. Considering the standard deviation of the light curve, we obtain an apparent opacity of $\rm{p}'= 0.08$ at the 1$\sigma$ level and $\rm{p}'= 0.24$ at the 3$\sigma$ level, which represents an apparent optical depth of $\tau' = 0.25$ (3$\sigma$). Here, we considered the full resolution of 8.65 km allowed by the data set.  

\begin{figure}[!h]
    \centering
    \includegraphics[width=\hsize]{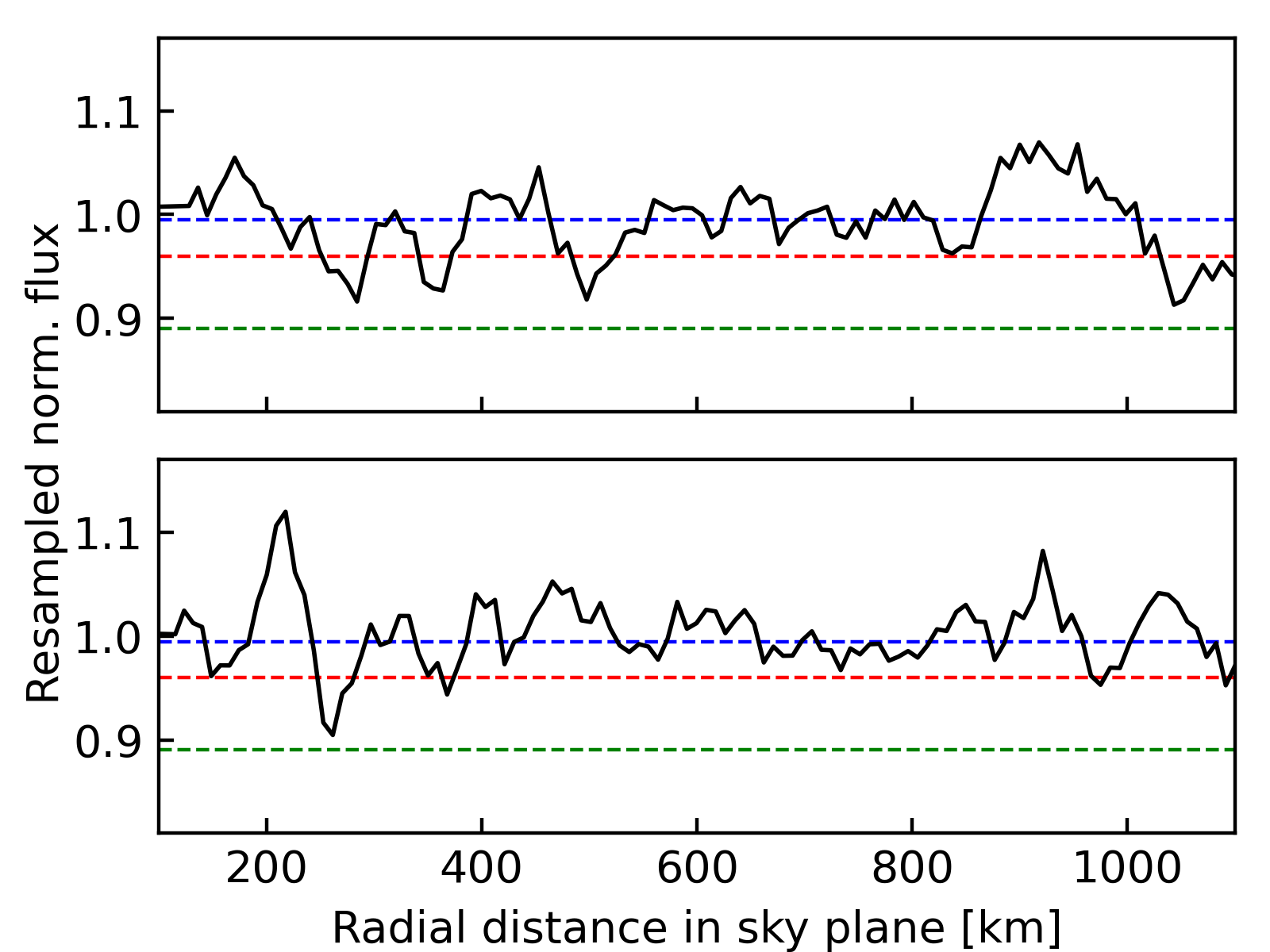}
    \caption{Similar as Fig. \ref{opacity2018}, but the data were smoothed to a resolution of 45 km per data point.}
    \label{shell45}
\end{figure}

\begin{figure}[!h]
    \centering
    \includegraphics[width=\hsize]{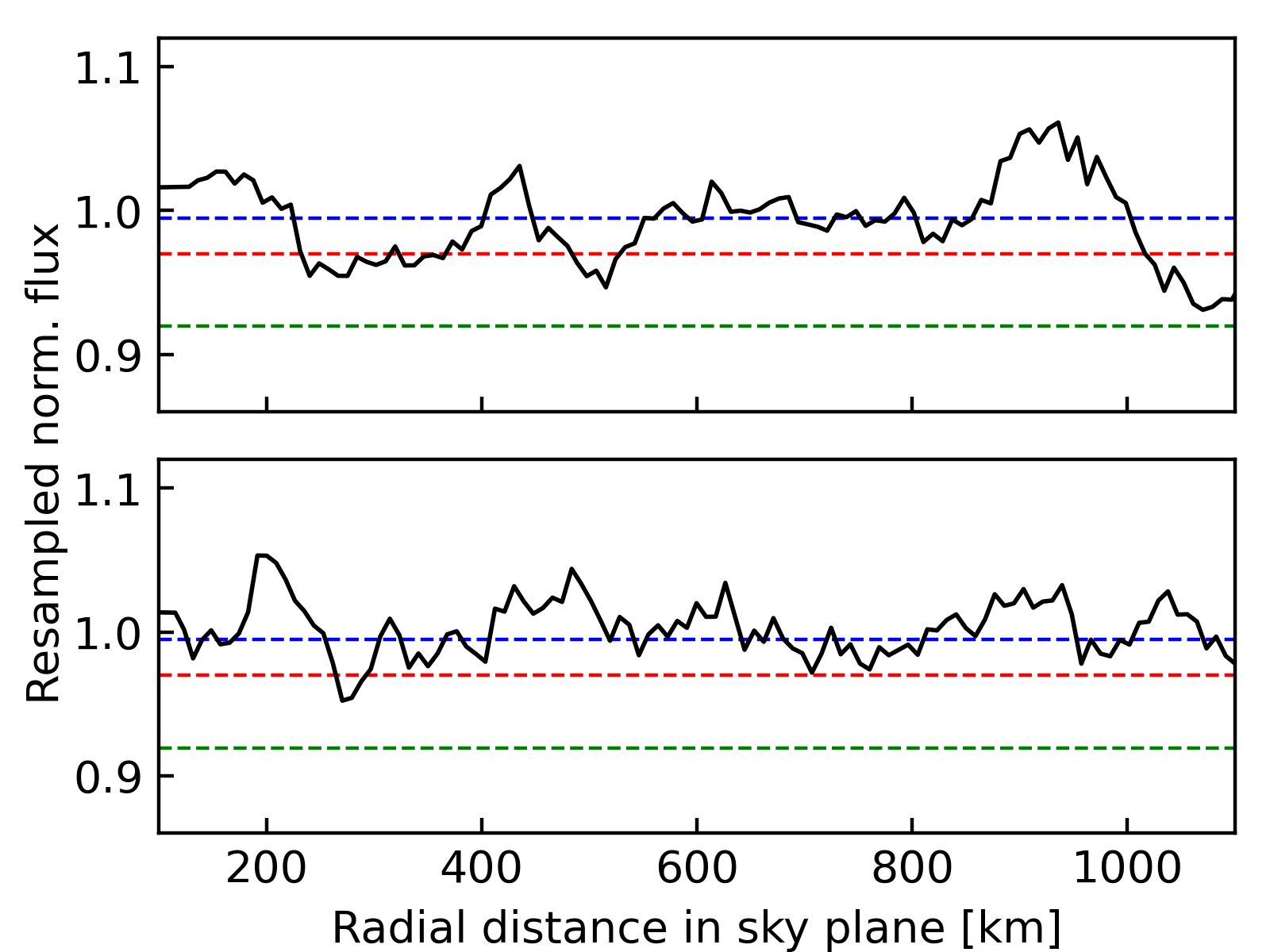}
    \caption{Similar as Fig. \ref{opacity2018}, but the data were smoothed to a resolution of 81 km per data point.}
    \label{shell81}
\end{figure}

If we consider that a shell of material can be spread over a large region, we should also search for symmetrical drops over several kilometers radially to the object's center. To do so, we binned the data using windows of 45 and 81 km \cite[according to the proposed structures in][]{elliot1995}
and obtained apparent optical depth limits of $\tau'=$~0.11 and 0.08 at the 3$\sigma$ level, respectively. As we can see in Figs. \ref{shell45} and \ref{shell81}, no symmetrical or smooth flux drops are seen in both graph.

\vspace{-0.5cm}

\section{Conclusions}

We have presented here the first multi-chord stellar occultation by Chiron. Sizes determinations were made using different techniques before this work.For instance,
\cite{Groussin2004} combined visible, infrared, and radio observations to derive a diameter of 142~$\pm$~10~km; \cite{Fornasier13} combined \textit{Herschel} and \textit{Spitzer} thermal data to derive (among other solutions) an equivalent diameter of 218~$\pm$~20~km.
\citet{Lellouch2017} reanalyzed the previous data adding observations made with ALMA to derive an equivalent diameter of 210~$\pm$~10~km for a spherical model. They have also considered an elliptical model and the influence of the putative rings, but they found that neither help to reconcile the discrepant diameter determinations. We decided to use their spherical model as the most conservative approach and give reliable limits to Chiron's apparent equatorial diameter (R$\rm _{equat}=126\pm22$~km) and oblateness ($\epsilon=0.27\pm0.18$).

Considering the derived true rotational light curve amplitude of $\Delta m_0=~$0.16~$\pm$~0.03 \citep{Groussin2004} and assuming a
Jacobi shape (i.e., fluid hydrostatic equilibrium), we can expect Chiron's semi-axis to be $a=126\pm22$~km,  $b=109\pm19$~km, and $c=68\pm13$~km, implying a model-dependent density of  $1,119~\pm~4~\rm{kg~m^{-3}}$, for a rotation period of 5.917813~h, and a volume-equivalent radius of R$_{vol}$= 98~$\pm$~17~km. With these values, we can obtain the first estimation of Chiron's mass of $4.8~\pm~2.3\times10^{18}$~kg. When Chiron is viewed equator-on, the average equivalent surface radius is R$_{Sequiv}=89\pm16$~km, it can be used with the true absolute magnitude of 7.28~$\pm$~0.08 in V band obtained by \citet{Groussin2004} and Mag${\rm{_V}}=~-26.74~$ for the Sun, to calculate Chiron's geometric albedo to be $p\rm{_V}~=~0.076\pm0.026$, which is not in agreement with their value of 0.11~$\pm$~0.02; however, as discussed by \citet{Ortiz2015}, they did not consider the presence of the putative rings, which may contribute to up to 30\% of the total reflected flux. Even if Chiron does not exhibit a Jacobi equilibrium shape, which might be the case, it is reasonable to consider that it does not diverge much from the here proposed shape.

We reanalyzed the secondary events observed in 2011 using Faulkes Telescope \citep{Ruprecht2015,Sickafoose2020} and obtained the properties of the rings using the same procedures as \citet{Braga-Ribas2014, berard2017}, considering the size and pole orientation proposed by \citet{Ortiz2015}.

\begin{table}
\caption{Chiron's astrometric positions from 2018 and 2019 occultations.}             
\label{astrometry}  
\centering       
\small
\begin{tabular}{c c c }      
\toprule  \toprule  
Event date     &  Right ascension      &     \multirow{2}{*}{Astrometric}  \\
(dd-mm-yyyy)                &  (hh mm ss.s)         &     \multirow{2}{*}{uncertainty}  \\
Instant (UT)                & Declination           &     \multirow{2}{*}{(mas)}        \\
(hh:mm:ss.s)                & ($^\circ$ ' '')       &   \\
\midrule    
28-11-2018                  & 23 46 04.35419        &   $\pm$ 0.26     \\
20:50:43.14        & +02 13 05.25822        &   $\pm$ 0.35 \\
\midrule
08-09-2019                  & 00 10 12.73280        &   $\pm$ 0.91     \\
23:04:26.20                 & +04 37 05.26207       &   $\pm$ 0.45     \\
\bottomrule
\end{tabular}
\end{table}

Data obtained from South Africa Astronomical Observatory allowed for a search for signatures of the proposed ring system. No clear evidence of the presence of the rings was found, mainly due to the long exposure time needed to record the occultation event. If the structures observed in 2011 were present in the 2018 data, they would have been detected at 2.6~$\sigma$ during the ingress and 2.2~$\sigma$ at the egress. These values are too low to claim any detection. Figure \ref{opacity2018} clearly shows that many drops are below the 1~$\sigma$ level, but none below the 3$\sigma$ level. We note that a ring such as the C1R, the main ring of (10199) Chariklo \citep{Morgado2021}, would have been detected above the 3~$\sigma$ level.

We also searched for broad features, such as a shell of material, and found no evidence. The upper limits are $\tau'=$~0.25, 0.11, and 0.08 for shells spread over $\sim$9~km, 45~km, and 81~km, respectively. %
\citet{elliot1995} reported a feature (F2) with width $\sim$74~km and a maximum optical depth of $\tau '\sim$0.11. Considering the detection limit for apparent optical depth obtained with the 81~km-smoothed curves, the F2 feature would be detected outside the 3-$\sigma$ level.

\begin{figure}[!ht]
    \centering
    \includegraphics[width=\hsize]{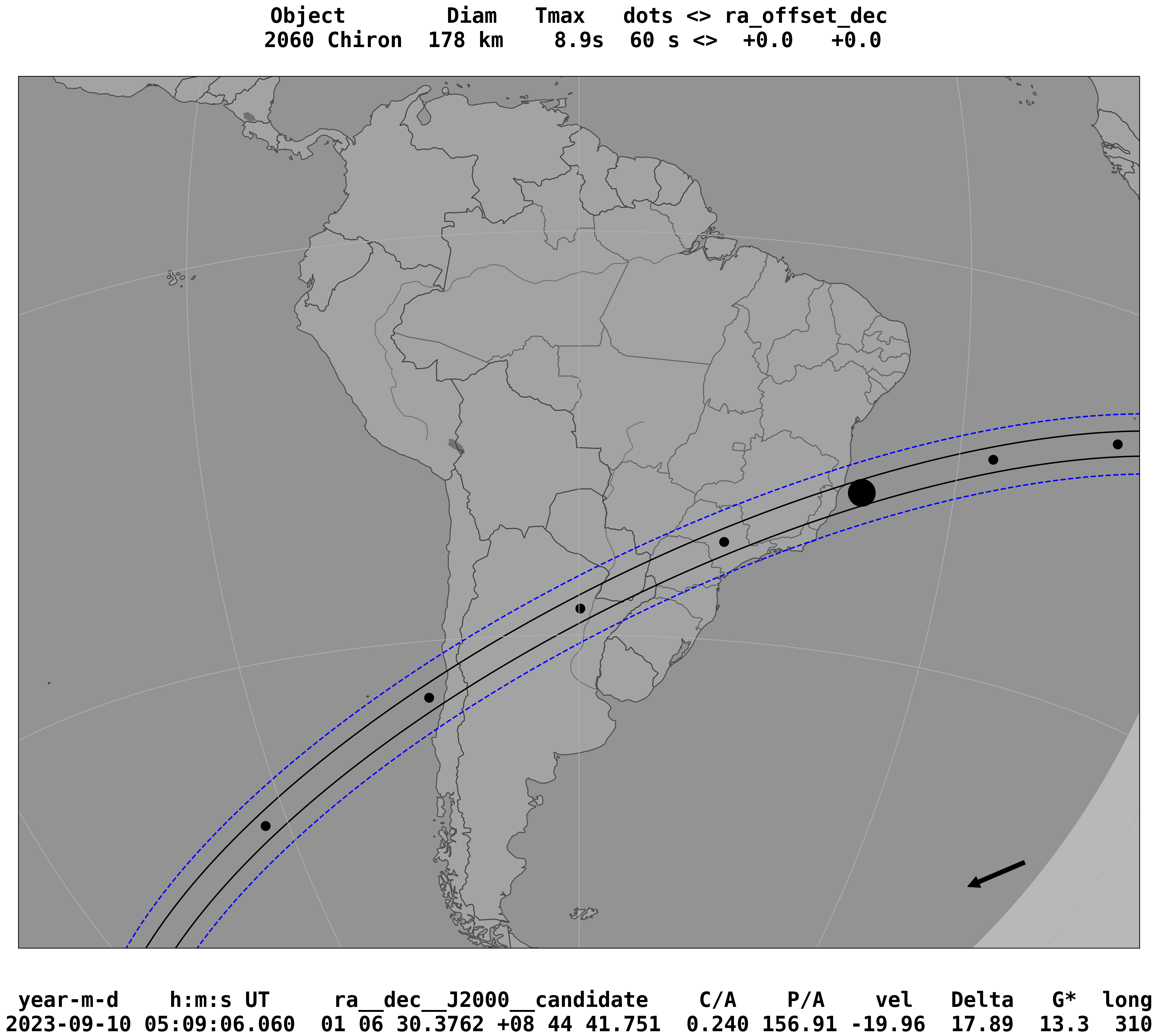}
    \vspace{-0.5cm}
    \caption{Most promising upcoming occultation event by Chiron on September 10, 2023. Solid black lines correspond to the main body shadow path, while the blue dashed lines the projected distance of the proposed rings. Prediction based on the latest NIMA ephemeris (NIMAv15, available on the Lucky Star web page), using astrometric positions from all the detected occultations by Chiron and the \Gaia DR3 catalog.}
    \label{futuremaps}
\end{figure}

Chiron is currently in a region of the sky characterized by a low density of stars, so relevant occultation events by this object are very rare. Thus, observations of new occultation events are needed. Adding the astrometric positions of the events presented here (Table \ref{astrometry}) and those presented in \citet{Rommel2020}, we searched for future events. A promising event will occur on September 10, 2023 (Fig. \ref{futuremaps})\footnote{Further information can be found in the Lucky Star event's webpage \url{https://lesia.obspm.fr/lucky-star/occ.php?p=123202}}, which may allow for more precise size and shape determinations and contribute further information on Chiron's environment.


\vspace{0.5 cm}
\begin{acknowledgements}

This work was carried out within the Lucky Star umbrella that agglomerates the efforts of the Paris, Granada, and Rio teams, which is funded by the European Research Council under the European Community H2020 (ERC Grant Agreement No. 669416). This study was financed in part by the Coordenação de Aperfeiçoamento de Pessoal de Nível Superior - Brasil (CAPES) - Finance Code 001 and the National Institute of Science and Technology of the e-Universe project (INCT do e-Universo, CNPq grant 465376/2014-2). The following authors acknowledge the respective CNPq grants: F.B-R 314772/2020-0; BEM 150612/2020-6; RV-M 307368/2021-1, 401903/2016-8; J.I.B.C. 308150/2016-3; M.A. 427700/2018-3, 310683/2017-3, 473002/2013-2. The following authors acknowledge the respective grants: C.L.P thanks the FAPERJ/DSC-10 (E26/204.141/2022); B.E.M. thanks the CAPES/Cofecub-394/2016-05 grant; M.A. acknowledges F.A.P.E.R.J. grant E-26/111.488/2013; ARGJr acknowledges F.A.P.E.S.P. grant 2018/11239-8;  Some of the results were based on observations taken at the 1.6 m telescope on Pico dos Dias Observatory of the National Laboratory of Astrophysics (L.N.A./Brazil). This research made use of SORA, a Python package for stellar occultations reduction and analysis, developed with the support of ERC Lucky Star and LIneA/Brazil. This work has made use of data from the European Space Agency (E.S.A.) mission {\it Gaia} (\url{https://www.cosmos.esa.int/gaia}), processed by the {\it Gaia} Data Processing and Analysis Consortium (D.P.A.C., \url{https://www.cosmos.esa.int/web/gaia/dpac/consortium}). J.L.O., P.S-S., N.M., and M.K. acknowledge financial support from the grant CEX2021-001131-S funded by MCIN/AEI/10.13039/501100011033 and they also acknowledge the financial support by the Spanish grant nos. AYA-2017-84637-R and PID2020-112789GB-I00 and the Proyectos de Excelencia de la Junta de Andalucía grant nos. 2012-FQM1776 and PY20-01309. 

\end{acknowledgements}


\bibliographystyle{aa}
\bibliography{references.bib} 

\newpage
\begin{appendix} 
\section{Occultation Light Curves}
The occultation light curves obtained from the here reported events are presented in Fig. \ref{fig:LC}. The first four light curves from the 2019 event allowed to constrain Chiron's size and shape. The last light curve from the 2018 event allowed searching for secondary events around it.
\begin{figure*}[h!]
\centering
\includegraphics[width=8cm]{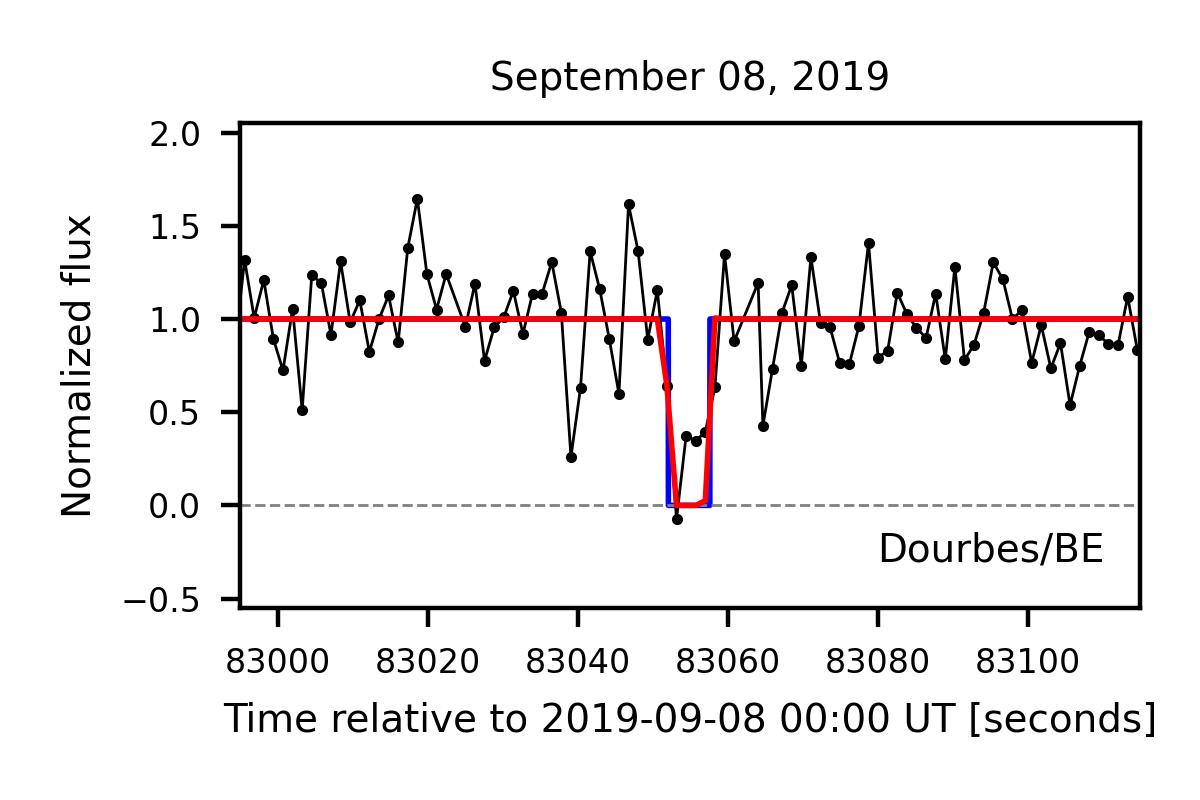}
\includegraphics[width=8cm]{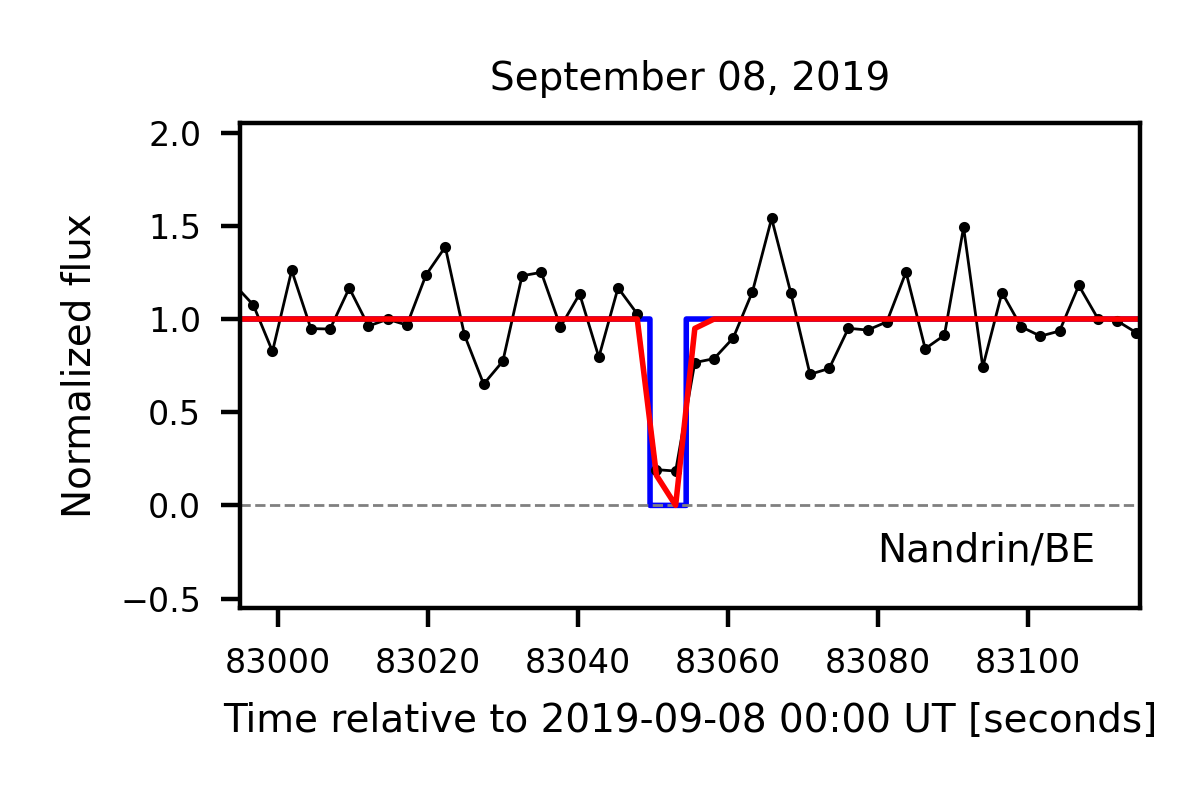}
\includegraphics[width=8cm]
{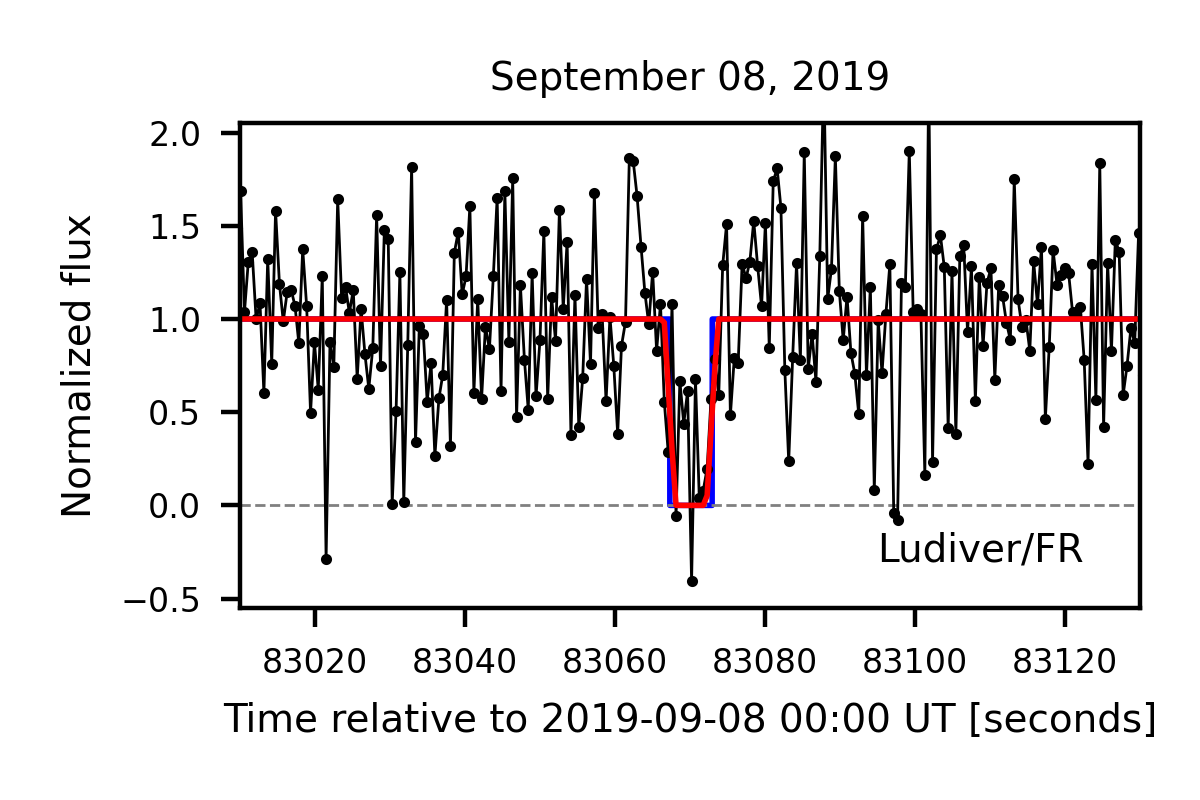}
\includegraphics[width=8cm]{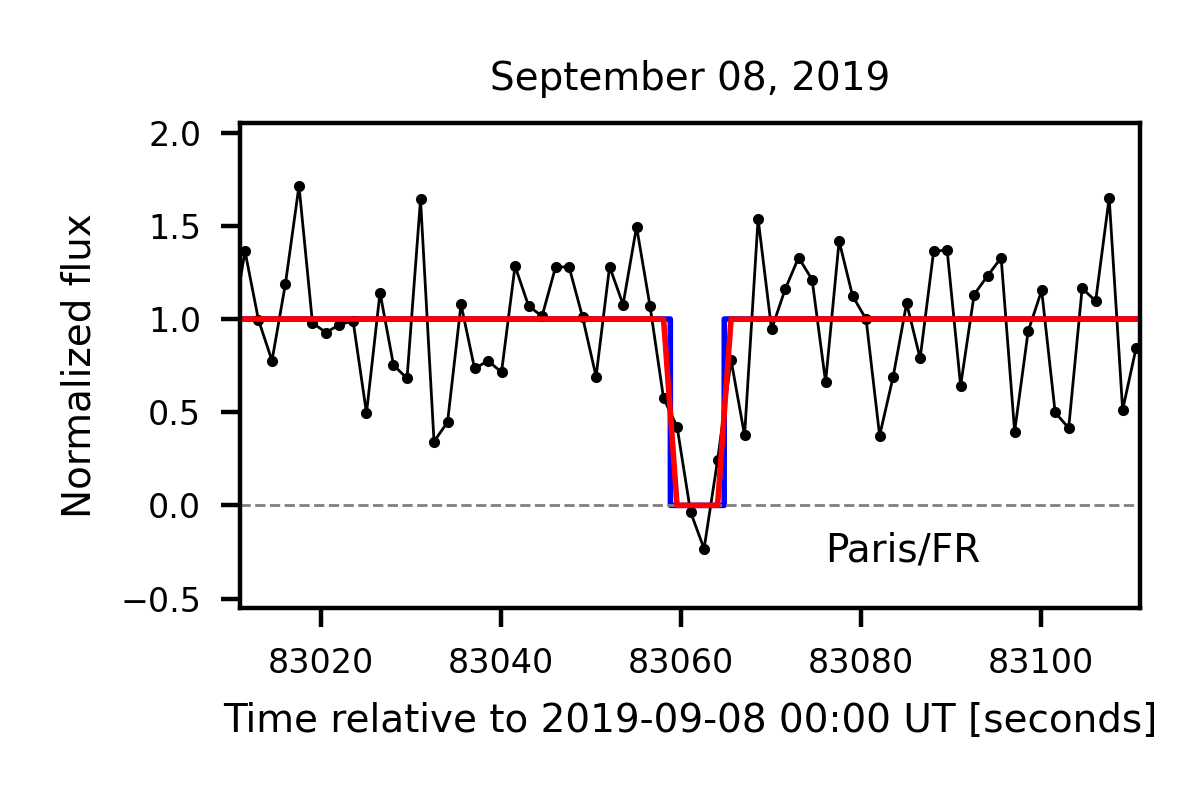}
\end{figure*}
\begin{figure*}
\centering
\includegraphics[width=8cm]{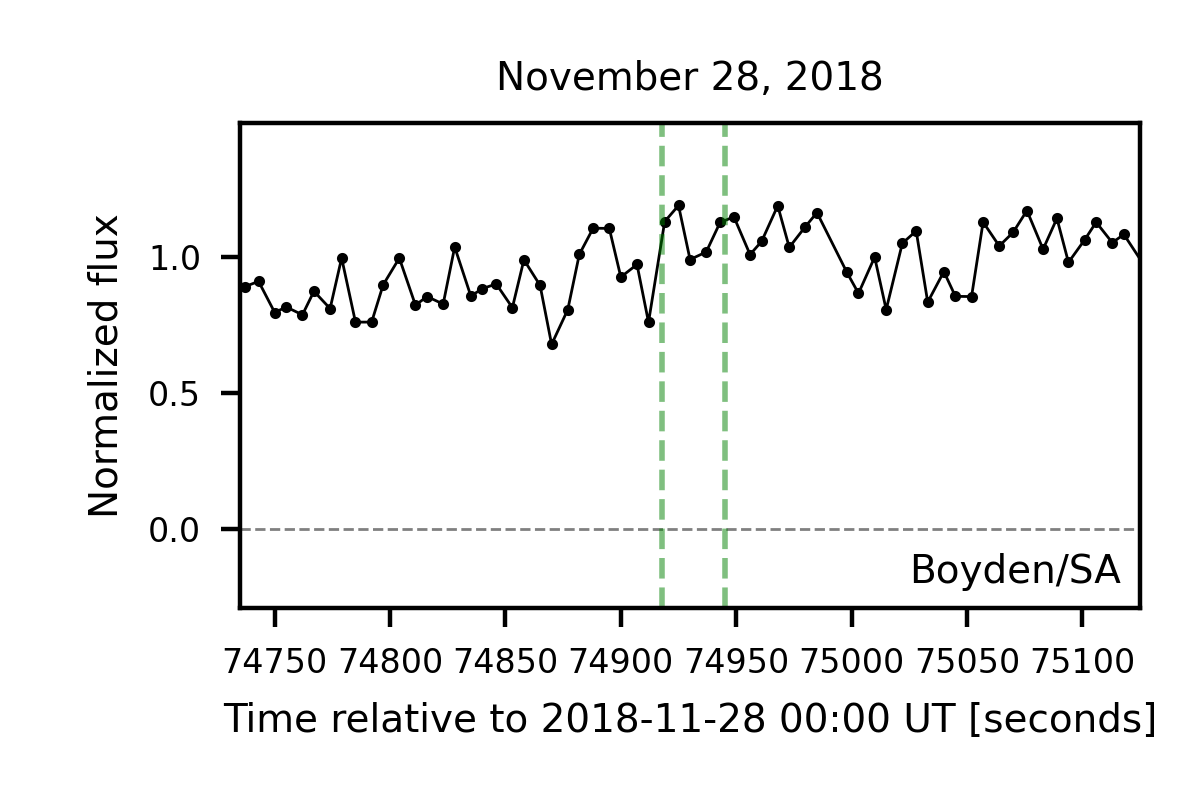}
\includegraphics[width=8cm]{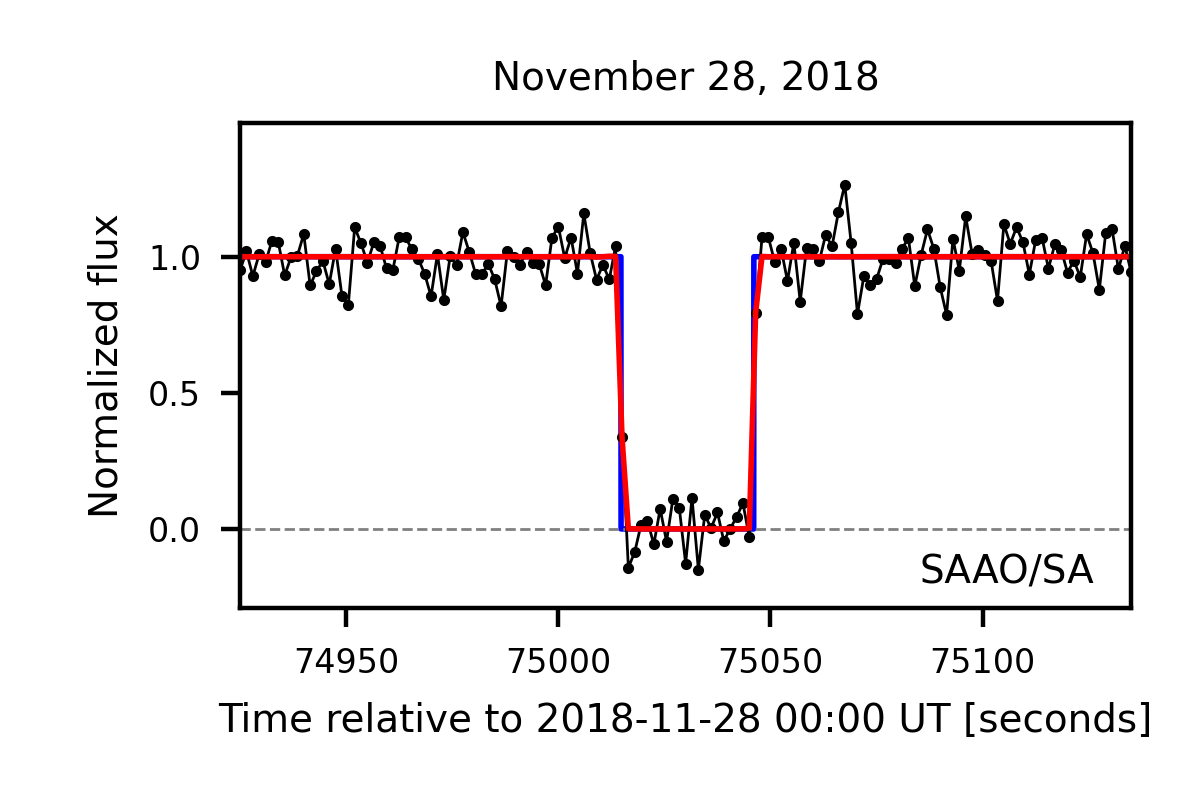}
\caption{Black lines and dots are the data obtained from each site, the blue lines the opaque band, and red lines the modeled light curve, convoluted to the exposure time. Times are given in seconds after midnight of the event date. The site names are given in each panel. The green vertical dashed lines on Boyden's panel indicate the theoretical ingress and egress times.}
\label{fig:LC}
\end{figure*}
\label{LCs}
\end{appendix}

\end{document}